\documentclass[preprint]{elsarticle}

\usepackage{ecrc}
\usepackage{url}
\usepackage{color}
\usepackage{xcolor}
\usepackage{amsthm,amsmath,amssymb,mathtools}
\usepackage{enumerate}
\usepackage{lineno}
\usepackage{tabularx}
\usepackage{booktabs}
\usepackage{float}

\newtheorem{theorem}{Theorem}

\usepackage{bbm}
\usepackage{subcaption}
\volume{00}

 \firstpage{1}

 \journalname{XXX Journal}

\runauth{}

\jid{procs}

\jnltitlelogo{Annual Review in Control}

\CopyrightLine{2011}{Published by Elsevier Ltd.}

\usepackage{amssymb}

\usepackage[figuresright]{rotating}

\begin{document}

\begin{frontmatter}




\title{Reinforcement Learning for Feedback-Enabled Cyber Resilience}


\author[label1]{Yunhan Huang, Linan Huang, Quanyan Zhu}

\address[label1]{Department of Electrical and Computer Engineering, New York University, 370 Jay Street, Brooklyn, New York, United States, 11201}

\begin{abstract}
Digitization and remote connectivity have enlarged the attack surface and made cyber systems more vulnerable. As attackers become increasingly sophisticated and resourceful,  mere reliance on traditional cyber protection, such as intrusion detection, firewalls, and encryption, is insufficient to secure the cyber systems. Cyber resilience provides a new security paradigm that complements inadequate protection with resilience mechanisms. A Cyber-Resilient Mechanism (CRM) adapts to the known or zero-day threats and uncertainties in real-time and strategically responds to them to maintain critical functions of the cyber systems in the event of successful attacks. Feedback architectures play a pivotal role in enabling the online sensing, reasoning, and actuation process of the CRM. Reinforcement Learning (RL) is an essential tool that epitomizes the feedback architectures for cyber resilience. It allows the CRM to provide sequential responses to attacks with limited or without prior knowledge of the environment and the attacker. In this work, we review the literature on RL for cyber resilience and discuss cyber resilience against three major types of vulnerabilities, i.e., posture-related, information-related, and human-related vulnerabilities. We introduce three application domains of CRMs: moving target defense, defensive cyber deception, and assistive human security technologies. The RL algorithms also have vulnerabilities themselves. We explain the three vulnerabilities of RL and present attack models where the attacker targets the information exchanged between the environment and the agent: the rewards, the state observations, and the action commands. We show that the attacker can trick the RL agent into learning a nefarious policy with minimum attacking effort. Lastly, we discuss the future challenges of RL for cyber security and resilience and emerging applications of RL-based CRMs.
\end{abstract}


\end{frontmatter}


\setcounter{secnumdepth}{3}
\setcounter{tocdepth}{3}
\tableofcontents


\section{Introduction} \label{sec:intro}
Recent attacks such as SolarWinds \cite{oxfordsolarwinds} and the ransomware attacks on the U.S. gas pipelines \cite{nicol2021ransomware} have shown intensifying concern of cyber threats on industrial and government cyber systems. These unprecedented attacks had created significant disruptions in business and government operations. We have witnessed not only a surge in the number of attacks but also their increasing sophistication. Now many attacks can circumvent traditional methods such as intrusion detection systems, firewalls, and encryptions. Advanced Persistent Threats (APTs) \cite{cole2012advanced} are one of such threats that are known for their stealthiness, intelligence, and persistence. 

Protection against such attacks becomes increasingly challenging. An attacker has to know only one vulnerability to launch a successful attack. In contrast, a defender needs to prepare for all its vulnerabilities and the associated attacks to secure the system successfully. 
{
Among those exploitable vulnerabilities, many of them are unknown until an attack exploit them (known as the zero-day vulnerabilities) or known but not patched timely (known as the $1$-day or $N$-day vulnerabilities due to the delay of $1$ or more days between the disclosure time and the attack time). An exploit directed at a zero-day vulnerability is called a zero-day attack,}  
which is exceptionally difficult, if not possible, for the defender to design a protection mechanism to prevent the system from. This information disadvantage for the defender has made perfect protection not only difficult to achieve but also cost-prohibitive. 

Furthermore, many recent attacks have targeted government agencies and critical infrastructures. The attackers are often supported by a nation-state or state-sponsored group. They are equipped with sophisticated tools, can conduct meticulous research about the system, and afford a prolonged period of persistent disguise. In contrast, security is often a secondary or add-on concern when designing or operating the system. A defender may not have invested sufficient resources in the protection mechanism. The striking disparity in the resources puts the defender at another disadvantage. 

Both information and resource disadvantages have made a mere reliance on cyber protection an insufficient way to safeguard our networks. There is a need to shift the focus from cyber protection to a new security paradigm. Cyber resilience offers such a perspective \cite{kott2019cyber,linkov2019fundamental}. It accepts the reality that the attacker can be successful in their attacks and complements the imperfect protection with resilience mechanisms. A cyber-resilient system adapts to the known and unknown threats and adversities in real-time and strategically responds to them to maintain the critical functions and performance in the event of successful attacks. 

\subsection{P2R2 Cyber-Resilient Mechanism}
A cyber-resilient mechanism (CRM) can be decomposed into four stages: Preparation, Protection, Response, and Recovery. We also call it a P2R2 CRM. The first stage of the CRM is preparation. At this stage, we assess the security risk of the cyber system and design appropriate security policies, such as the deployment of honeypots or deception mechanisms \cite{al2019autonomous,pawlick2021game}, training of the employees \cite{greitzer2008combating}, proper configurations of detection systems \cite{zhu2009dynamic}, and the preparation of the backups and contingencies for recovery plans \cite{huang2018distributed}. Preparation is often done offline and ahead of the real-time operations. Good preparation can help facilitate effective prevention and fast response to unanticipated scenarios in later stages.

The second stage is prevention. At this stage, we implement the designed security policies and protect the cyber systems. Some attacks can be easily detected and thwarted in real-time because of the meticulous preparation and design of the security policies. For example, consolidating moving targeting defense \cite{zhu2013game} into the communication protocols would make it harder for the attacker to map out the traffic patterns and consequently thwart the denial of service attack. However, despite this effort, there is still a probability for an attacker to become successful, especially for a highly resourceful and stealthy one. 

\begin{figure}[H]
\vspace{-0mm}\centerline{\includegraphics[width=0.7\linewidth]{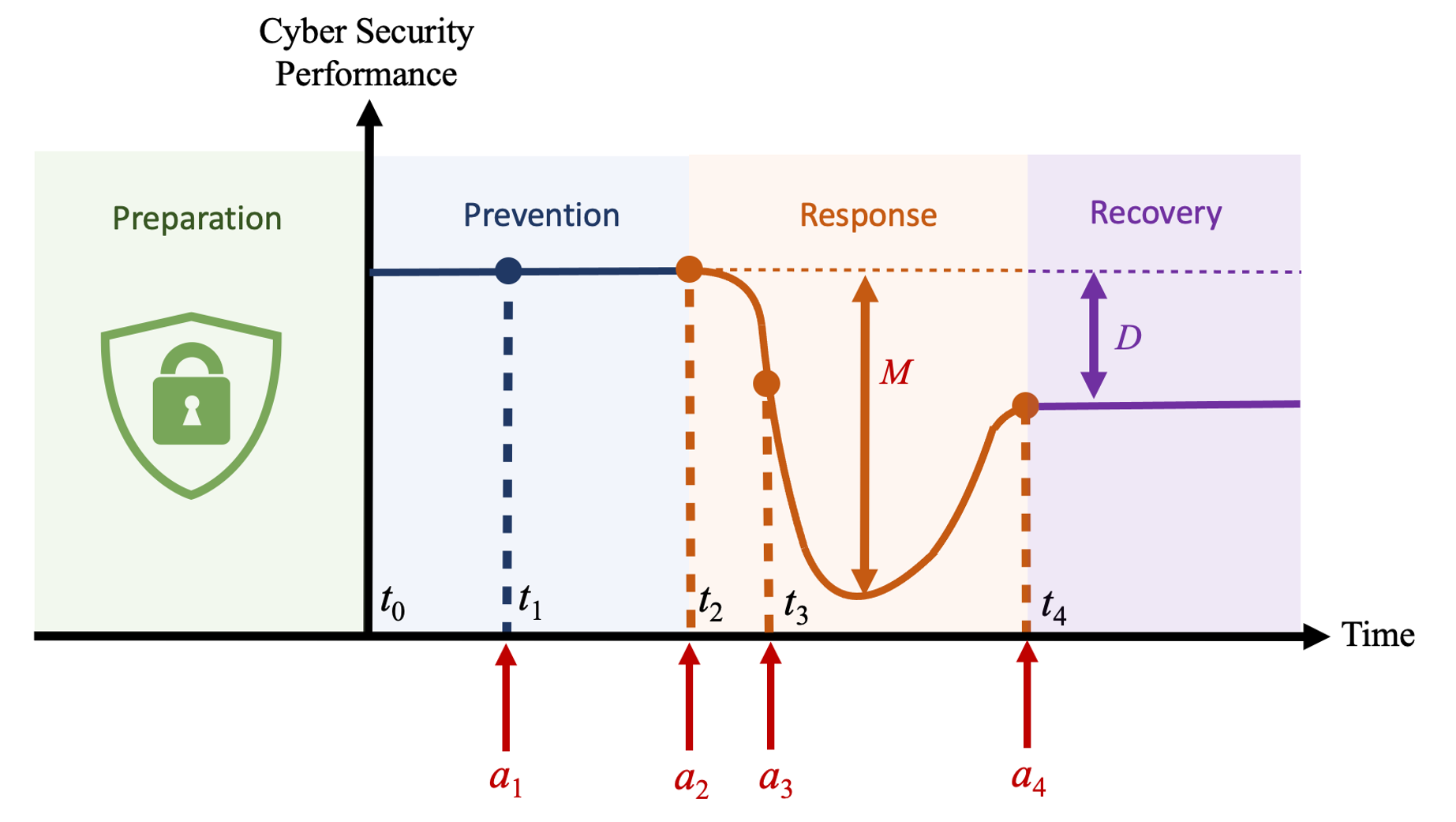}}\vspace{-2mm}
\caption{The four stages of a cyber-resilient mechanism: Preparation, Prevention, Response, and Recovery. An attacker launches a sequence of attacks at $t_1, t_2, t_3, t_4$. The cyber security performance decreases when an attack successfully penetrates the system at $t_2$. The system partially restores its performance at $t_4$.}\vspace{-4mm}
\label{crm}
\end{figure}

The third stage of response is critical to defending against attacks when we fail to thwart them at the prevention stage. At this stage, we acquire the information of the footprint of the attacker and reconfigure the cyber system to minimize the further risk of the attack on the cyber system \cite{reiger2019industrial,haque2018cyber}. The response to attacks is delay-sensitive. The information acquisition and the reaction to the observables need to be fast and done before the attacker moves onto the next stage of its attack. A fast response mechanism would rely on the acquired information and the designs at the preparation stage. There would be a cost in usability and performance when we respond to the attacks. The optimal response mechanism is a result of studying such tradeoffs.

The fourth stage of the CRM is recovery. The goal of the recovery stage is to reduce the spill-over impact of an attack and restore the cyber system performance as much as possible. The response to attacks in real-time is often at the sacrifice of the performance of the cyber system. There is a need to maintain system's operation and gradually restore its functionality to normal while reacting to the attacks.

Fig. \ref{crm} illustrates the four stages of a typical CRM over a timeline. Before the operation, the cyber system designs a protection mechanism by considering the anticipated and known attacks and the risks of its own system. The cyber system starts to operate at $t_0$. At time $t_1$, an attacker launches attack $a_1$. As the cyber system prepares for it, this attack is successfully thwarted, and the system is protected until the attacker launches a new attack $a_2$ at time $t_2$. The cyber system did not prepare for this attack. The attack accomplishes its goal (e.g., lateral movement or penetration of the system) and, as a result, the cyber risk increases or the cybersecurity performance drops. The cyber system needs to respond to the successful attack and the ensuing attack $a_3$ at time $t_3$. After fast learning from attacker's footprint, the cyber system makes strategic decisions to reconfigure and adapt to the adversarial environment and gradually improves the security posture of the system. The adapted system becomes more robust to attacks. The ensuing attack $a_4$ at time $t_4$ can be thwarted by the adapted protection mechanism. The cyber system restores to its best-effort post-attack performance at $t_4$.

From Fig. \ref{crm}, one can measure the cyber resilience across the four stages by the time it takes from the first attack to the recovery, i.e., $T=t_4-t_2$. This time interval indicates how fast the response is \cite{zhu2015hierarchical,zhu2020control}. The worst-case performance degradation between the interval $t_2$ and $t_4$ is $M$. The gap between the restored performance and the initial or planned performance, i.e., $D$, shows the effectiveness of the resilience.  The natural goal of the CRM is to minimize $T, M,$ and $D$. It is also evident that these metrics not only depend on the cyber system itself but also the type of attacks that the adversary launches. 
An optimal design would require understanding what threats are preventable and what can be mitigated by resilience.  

\subsection{Feedback Architecture and Reinforcement Learning}
\begin{figure}[h]
\vspace{-0mm}\centerline{\includegraphics[width=0.6\linewidth]{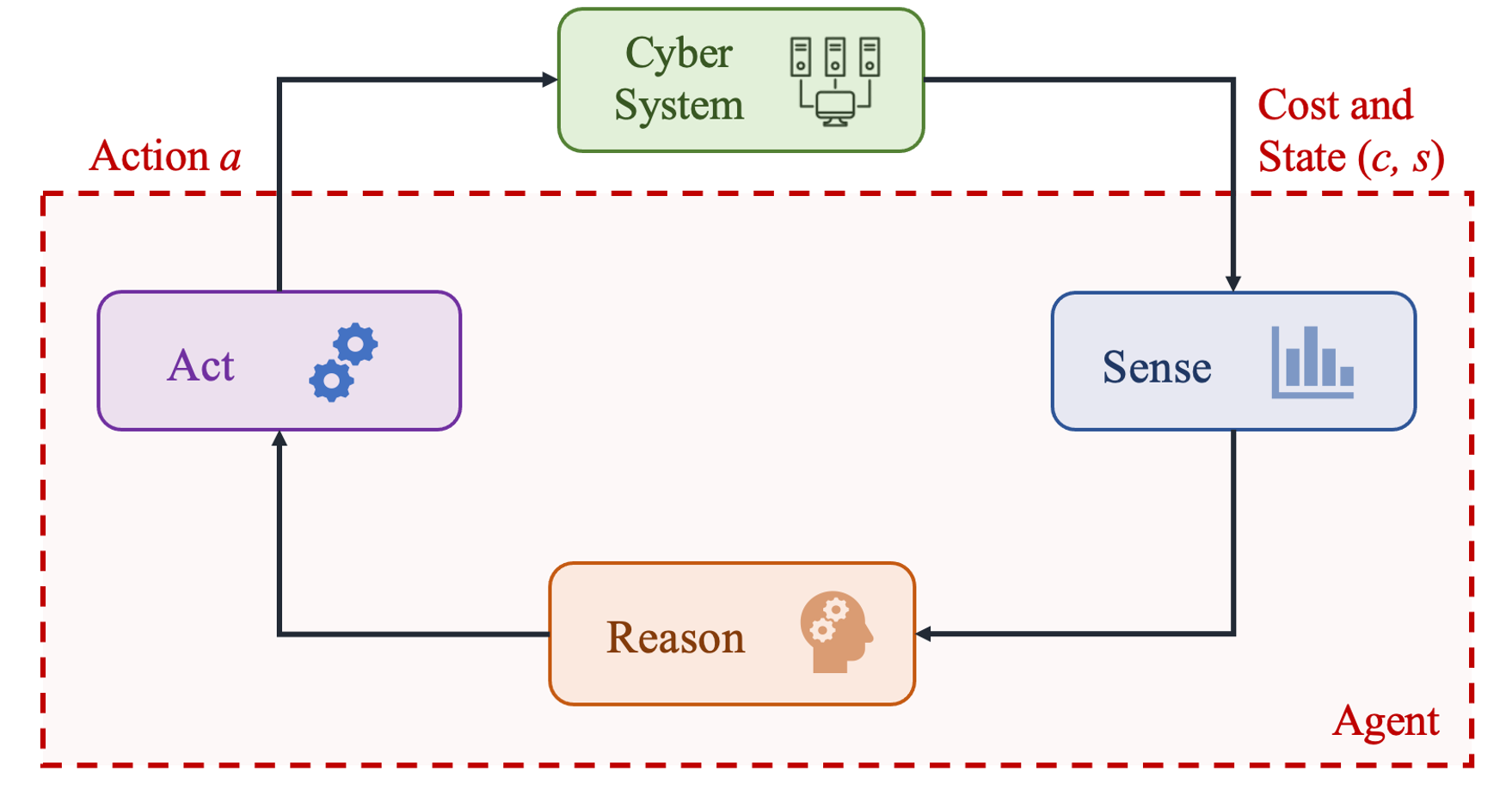}}\vspace{-2mm}
\caption{Feedback structure that enables cyber resilience. At each iteration, the cyber-resilient mechanism senses the parameters (e.g., cost and state) of the cyber system and makes decisions to act on the cyber system to reduce the cyber risks.}\vspace{-4mm}
\label{feedback}
\end{figure}

One critical component of the P2R2 CRM for cyber resilience is the response mechanism, which creates a dynamic process that involves information acquisition, online decision-making, and security reconfiguration. These components are strongly interdependent. One way to characterize this interdependence is through a feedback loop, as illustrated in Fig. \ref{feedback}. 
The feedback structure provides a system and control approach to cyber resilience \cite{jajodia2020adaptive}. The cyber system can be viewed as a plant. The CRM collects information from the cyber system through sensing and observation. The acquired information is used to reason about the current situation and decide how the system should react. The CRM acts on the recommended decisions by reconfiguring the security parameters and updating the cyber systems. The reasoning for the optimal cyber response can be viewed as a controller. The security reconfiguration can be interpreted as the actuation of the cyber system. The control perspective provides a rich set of tools and concepts from control theory to design an optimal, robust, and adaptive CRM to improve cyber resilience.

One pivotal control framework to {achieve an optimal, robust, and adaptive CRM} is reinforcement learning (RL). Similar to classic control approaches, RL leverages the idea of feedback architecture to guarantee certain aspects of performance such as optimality and stability. However, RL approaches are distinguished from classic control approaches from several facets.

First, RL allows the CRM policies to adapt to the online observations without knowing the underlying dynamics. The CRM can only observe the state $s$ and the cost $c$ from the cyber system at each step and then choose action $a$ to optimally improve cyber resilience. RL is particularly useful for cyber systems as their precise models are often unavailable, and the influence of the attacker makes it even harder to map out an accurate cyber system model. There are also many unknowns in the attack model and cyber systems that require online learning and adaptation. RL provides an appropriate framework to bridge the gap between control theory and cyber resilience. 

{Second, RL allows the CRM to have a more generic space that captures the activities in the cyber system. Classic control oftentimes deals with physical systems such as mechanical systems \cite{de2000lyapunov}, electrical circuits \cite{tsubone1998stabilizing}, chemical systems \cite{skogestad2004control} and biological systems \cite{huang2019differential}. The state space of such systems is mostly continuous and finite dimension represented by $\mathbb{R}^n$ with $n$ being a finite integer. Cyber systems are computer \& communication systems integrated with physical systems whose state space is discrete or hybrid (discrete and continuous) in most cases. RL is versatile at dealing with different types of state space. This versatility enables CRM to capture the complex status of cyber systems. Besides, RL equipped with function approximation techniques copes well with large-scale cyber systems that contain a large number of elements or have high-dimension state spaces \cite{malialis2013large,xu2007defending}.
}

{
Third, RL requires different stages than the classic control approaches before being implemented in real systems. Specifically, the implementation of RL techniques undergoes three phases: the training phase, the testing phase, and the execution phase. In the training phase, the agent interacts with the environment or the simulator to obtain data and improves his/her policy using the obtained data. In the testing phase, we evaluate the performance of the improved policy. If the policy reaches optimality or a satisfactory level of sub-optimality, the learned policy (or the learned CRM in our case) will be put into use, and the implementation enters the execution phase. Classic control approaches go through four phases: the modeling phase, the design phase, the testing phase, and the execution phase. In the modeling phase, modelers with domain knowledge build a mathematical model for the underlying system. In classic control, the training phase is replaced by a design phase where experts with domain expertise design a policy/controller based on the model constructed for the cyber system. The designed policy is then tested and evaluated before being executed in real systems. The training phases consume an enormous amount of data to improve the learned CRM constantly. Much effort need to be spent obtaining the data through either simulation or interacting with the environment. Data-inefficiency needs to solved by combining modern RL techniques with domain expertise from cyber security experts.  }

\subsection{System Vulnerabilities and RL-enabled Defense}
The uncertainties in the cyber system are caused by not only the complexity in its configuration but also the incomplete information of the attack models. An attack model specifies the vulnerabilities an attacker would exploit on the attack surface of the cyber system. The design of the CRM pivots on the type of vulnerabilities that the defender aims provide resilience for if he cannot succeed in defending against attackers who exploit them. We categorize the vulnerabilities into three major types. The first one is the class of posture-related vulnerabilities that arise from the resource disadvantage of a defender. The defender is not able to prepare for all vulnerabilities on the attack surface and has to rely on CRM to overcome this disadvantage. The second one is the class of information-related vulnerabilities that result from the information asymmetry between an attacker and a defender. The attacker has more information about the defender than the defender has about the attacker. The CRM is useful to tilt the information asymmetry and allow the defender to create uncertainties for the attacker while gaining reconnaissance of the attacker. The third type of vulnerability is human-induced. Humans are the weakest link in cyber protection. Social engineering, human errors, and insider threats are common attack vectors that an attacker can use to gain access to cyber systems before further penetrating the systems and reaching the targeted assets. The CRM can monitor human performance and guide humans to reduce the risks of their behaviors. 



We use three applications to elaborate the design of CRMs for the three major types of vulnerabilities. 
The first application designs moving target defense (MTD) for posture-related vulnerabilities that result from the defender's natural disadvantages. The MTD creates reconnaissance difficulties for attackers by introducing uncertainties into the cyber systems. 
{
The authors in  \cite{zhu2013game} explore the capability of RL in changing the system configurations to create the most uncertainty for the attackers while minimizing the change cost, which leads to a trade-off between security and usability. Both attackers and defenders apply RL to learn the risk and update their policies.} 

The second application addresses the information-related vulnerabilities that arise from the defender's information disadvantage. Deception technologies, including honeypots and honeyfiles, have been used to reduce information asymmetry between attackers and defenders. 
{
The authors in  \cite{huang2019adaptive} use RL to design the deception technologies to engage the attacker and obtain threat information. The goal is to extract the most threat intelligence from attacks while minimizing the risks of detection and evasion. The study leads to} the security insight that compared to an immediate ejection after detection, it can be more beneficial to engage attackers in the honeypots to obtain threat intelligence. 

The third application designs CRM to mitigate human-related vulnerabilities. 
{
Adversaries have exploited human inattention to launch social engineering and phishing attacks toward employees and users. 
Compared to the \textit{reactive} attentional attacks that exploit the existing human attention patterns, \textit{proactive} attentional attacks can strategically change the attention pattern of a human operator or a network administrator. 
The authors in \cite{RN661,huang2021radams} have identified a new type of attacks called Informational Denial-of-Service (IDoS) attacks that generate a large volume of feint attacks to overload human operators and hide actual attacks among feints. 
They create an RL-based CRM to learn the optimal alert and attention management strategy.}  

\subsection{Reinforcement Learning in Adversarial Environment and Countermeasures}

As we use RL to improve the resilience of the cyber system, an intelligent attacker can also seek to compromise or mislead the RL process so that the cyber system ends with a worse or vulnerable security configuration. In this case, RL enlarges the attack surface and creates opportunities for the attacker. Hence it is essential to safeguard the RL while using it to secure the cyber system. Section \ref{sec:RLAdvEnv} presents several attack models that target at RL algorithms and discusses their impact and ways to secure them. RL agents learn a `good' policy through interacting with the environment and exchanging key information such as the rewards, the state observations, and the control commands. The three exchanged signals, {which can be exploited by the attacker}, become the vulnerabilities of most RL-enable systems. To understand RL in an adversarial environment, the first is to understand the adversarial behaviors of the attackers. To do so, we need to craft the attack models, which specify the attacker's objective, the attacks available to the attacker, and the knowledge the attacker has. The second is to understand how {different types of attacks} affect the learning results of RL algorithms, which requires the assessment of the performance degradation of the {RL-enabled} systems. With the understanding of {attack models} and their impact on the learning results, we can develop countermeasures to mitigate the effect of the attacks and to protect RL-enabled systems from attacks.

We review the {current studies on security of RL} and {focus on attacks on the three signals}, i.e., attacks on the rewards, attacks on the sensors, and attacks on the actuators. We {present a sequence of} attack models that can achieve a significant impact on RL-enabled systems with only a tiny effort on attacking. Given that this branch of research is still in its infancy, we identify several research gaps that researchers from the learning community and the control community can help narrow. One missing component is that most works focus on deceptive attacks that falsify the values of the feedback information. In contrast, few works study the DoS/jamming attacks on RL algorithms. Another missing component is the study of attacks on actuators and the defensive mechanisms against these attacks.

\subsection{Notations and Organization of the Paper}

Throughout the paper, we use calligraphic letter $\mathcal{X}$ to define a set and boldface letter $\mathbf{X}$ to denote a vector. 
Notation $\mathbb{E}$ represent expectation and $\mathbb{R}_+$ represents the set of positive real numbers. 
{
The notations $\mathbb{Z}_{\geq 0}$ and $\mathbb{Z}_{>0}$ represent the set of non-negative and positive integers, respectively.}
The indicator function $\mathbf{1}_{\{x=y\}}$ equals one if $x = y$, and zero if $x\neq y$. If two sets $B\subset A$, $A\backslash B$ denotes the set $\left\{x\ \middle\vert x\in A, x\notin B \right\}$. 
If $\mathcal{A}$ is a finite set, then we let $\Delta \mathcal{A}$ represent the set of probability distributions over $\mathcal{A}$, i.e., $\Delta \mathcal{A}:=\{p:A \leftarrow  \mathbb{R}_{+} | \sum_{a\in \mathcal{A}} p(a)=1\}$. {The main notation of this paper is introduced in Section \ref{subsec:RLalgos}. For sections that are dense with parameters, we include separate notation tables to help readers navigate.}

The remainder of the paper is organized as follows. 
{
    Section \ref{sec:RL} first introduces a general schema of RL methods to help readers navigate in the rich universe of RL algorithms. After that, Section \ref{sec:RL} gives an overview of literature that studies how RL is leveraged to fight against different types of attacks on cyber systems.
   While Section \ref{sec:RL} focuses on the different attacks, Section \ref{sec:reviewRLCR} centers on three major classes of vulnerabilities: the posture-related vulnerabilities, the information-related vulnerabilities, and the human-related vulnerabilities. In Section \ref{sec:reviewRLCR}, we discuss how each vulnerability can be addressed by RL.
   Following the discussion of \ref{sec:reviewRLCR}, Section \ref{sec:CRMechanismAndApplications} delves into three application domains, each representing one class of vulnerabilities, and presents design methodologies for cyber resilience.
    Compared to previous sections that are about RL for cyber security, Section \ref{sec:RLAdvEnv} is about security of RL, which introduces potential security threats faced by RL itself. In Section \ref{sec:RLAdvEnv}, we discuss the vulnerabilities of RL algorithms and provides an overarching review on this emerging research topic.
   We conclude the paper in Section \ref{sec:conclusions} and discuss several directions for future research.
}






\section{{Reinforcement Learning and Its Applications in Cyber Resilience and Defense}}\label{sec:RL}


The massive scale of Internet-connected or networked-connected systems expands the attack surface. The attacker can leverage one vulnerability in massive systems to penetrate the system and launch successful attacks. Most Internet-connected systems have dynamic nature due to their movement or changes in the environment. Hence, attacking windows can appear from time to time depending on the underlying dynamics, encouraging the attackers to look for opportunities persistently. The massiveness and the dynamics of the system require protecting and securing mechanisms to be responsive, adaptive, and scalable.

With the recent advances in Artificial Intelligence (AI), both the defense and the attacking sides have started to use AI techniques, especially machine learning (ML) methods, to obtain an edge. Sophisticated attackers leverage ML techniques to locate vulnerabilities, stay stealthy, and maximize the effect of attacks. Defenders employ
ML techniques to adaptively prevent and minimize the impacts or damages. Both supervised and unsupervised learning methods have been used widely by the defenders for spam filtering \cite{crawford2015survey},  intrusion detection \cite{buczak2015survey,sommer2010outside}, zero-day vulnerability forecasting \cite{last2016forecasting}, and malware detection \cite{xiao2018iot}. However, these traditional ML methods cannot provide dynamic and sequential responses against cyber attacks from the dark side with unknown patterns and constantly evolving behaviors. 

As a branch of ML, Reinforcement Learning (RL) is a learning paradigm that learns from its own experience over time through exploring and exploiting the unknown and changing environment. 
RL provides suitable tools for defenders to take sequential actions optimally without or with limited prior knowledge of the environment and the attacker. RL approaches allow the defender to capture various types of dynamics (e.g., stochastic or deterministic), a wide range of protective and defensive actions ( high-dimensional continuous state space or discrete state space), and different kinds of system states. RL demonstrates an excellent fit for cyberspace where cyber attacks become increasingly advanced, swift, and omnipresent \cite{ni2019multistage,uprety2020reinforcement,nguyen2019deep}. In recent years, the development of deep learning has fueled the successful synthesis of Deep Reinforcement Learning (DRL). The idea of DRL is to utilize a neural network to approximate complicated functions with high-dimensional inputs. The integration of deep learning enhances the conventional RL methods to capture the massive scale of many Internet-connected systems such as mobile networks and IoT systems \cite{zhu2018deep,ferdowsi2018robust}.


\subsection{RL Framework and Algorithms} \label{subsec:RLalgos}
The RL agent learns a near-optimal, if not optimal, to protect, prevent, or recover from attacks by constantly exploring and exploiting the environment in a feedback manner, as is shown in Fig. \ref{feedback}. In Fig. \ref{feedback}, $s$ represents the security state of the cyber system, and the cost $c$ is the monetary cost or performance degradation of the system induced by attacks and operations. The RL agent, or the defender, aims to find an action $a$ that optimally modifies the system configuration or other parameters, leading to changes in the security state and the cost. The feedback loop forms an iterative process of agent-environment interactions. 

RL is underpinned by a mathematical framework called Markov Decision Process (MDP). An MDP is denoted by a $5$-tuple $\langle\mathcal{S},\mathcal{A},c,\mathcal{P},\beta\rangle$, where $\mathcal{S}$ is the state space containing all possible states of the cyber system. The action space $\mathcal{A}$ denotes the actions available for the defender to protect the cyber system, to recover from the damage caused by attacks, or to mitigate the effect of attacks. The cost $c$ depends on the current and/or the next security states, and the current action, which is usually denoted by a function that maps $\mathcal{S}\times \mathcal{A}\times \mathcal{S} \rightarrow \mathbb{R}_+$. The transition kernel $\mathcal{P}$ defines the rule of how the system state evolves based on the actions taken by the defender. The discount factor $\beta$ is a weighting factor that assigns more weight to current costs than future costs. The goal of an MDP problem is to find a policy $\pi_t:\mathcal{S}\rightarrow \mathcal{A}$ to minimize a certain form of the accumulative costs $\sum_{t=0}^\infty \beta^t c_t$ over time. 

MDPs are generic modeling tools that can model the dynamic and feedback nature of various types of cyber systems. Conventional MDP approaches require the knowledge of the transition probability of the cyber system and the explicit definition of the reward function to find an appropriate strategy. However, in reality, obtaining such information is prohibitive for two reasons. The first is due to the complexity of the cyber system and its dynamics. The second is because the attackers are usually on the dark side whose behavior is unknown to the defender. RL solves the MDP problem without the knowledge of the transition kernel $\mathcal{P}$ and the cost function $c$. Instead of utilizing a prior known transition kernel and cost function, RL agent learns the optimal policy from sequences of states $\{s_t\}_{t\in \mathbb{Z}}$, costs $\{c_t\}_{t\in\mathbb{Z}}$, and actions $\{a_t\}_{t\in\mathbb{Z}}$, which are obtained by interacting with the environment over time.

\begin{figure}[H]
\vspace{-0mm}\centerline{\includegraphics[width=0.8\linewidth]{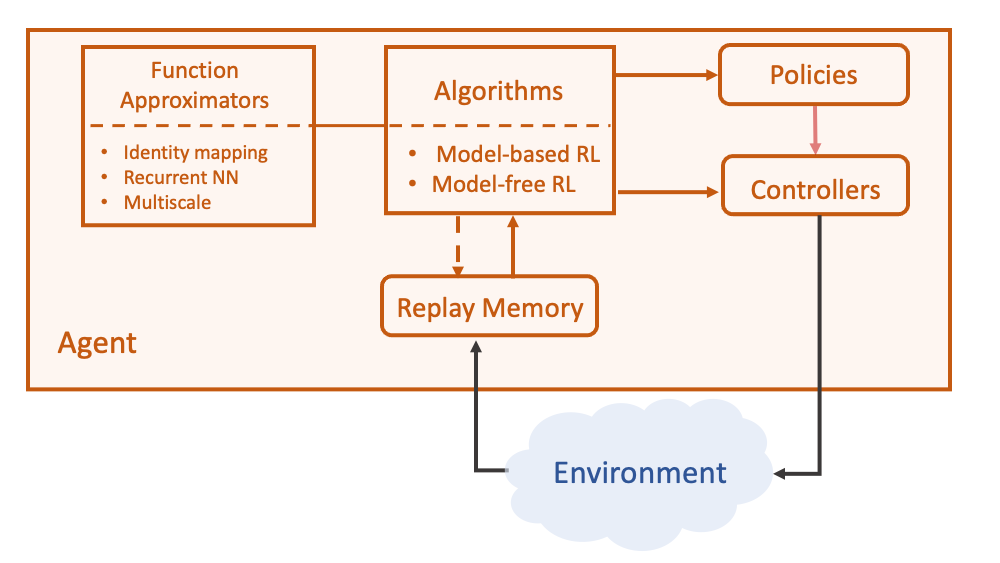}}\vspace{-2mm}
\caption{A general schema of RL algorithms.}\vspace{-4mm}
\label{fig:generalschma}
\end{figure}

{Having been actively studied for decades, RL has a rich universe of algorithms that help the agent find a satisfactory policy. Covering every single algorithm is beyond the scope of this paper. Here, we highlight a general schema of RL models shown in Fig. \ref{fig:generalschma} and present a concise taxonomy of RL algorithms.}

{{
\subsubsection{Model-Based and Model-Free}
One advantage of RL over classic MDP and control is its ability to cope with the unknown environment (e.g., cost function $c$, dynamics $\mathcal{P}$). Its ability to tackle unknown environments makes RL an appealing tool for cyber resilience. There are two types of methods that tackle the curse of modeling: the model-based RL and the model-free RL. Here, a `model' means an ensemble of acquired environmental knowledge. If the model (the cost function $c$ and the transition kernel $\mathcal{P}$) is known, methods in classic MDP (e.g., value iteration, policy iteration, linear programming, etc.) can be applied directly \cite{bertsekas1996neuro}. 

In model-based RL, the agent estimates the cost function as well as the transition kernel using data sequences $\{s_t,c_t,a_t,s_{t+1}\}_{t\in\mathbb{Z}}$. Once elements of the model are constructed/estimated from samples, the agent can apply classic methods in MDP directly to obtain a near-optimal policy. There are occasions where the model is explicitly given and can be accessed directly by the agent. For example, the MCTS (AlphaGo/AlphaZero) algorithm relies on a fully known environment because the rules of the GO game are exactly known \cite{silver2016mastering}. Nevertheless, in most cases, the model is unknown due to the complexity or opaqueness of the environment. The agent estimates the elements in the environment first, then applies classic MDP methods. Typical examples of this kind includes the MBMF algorithm \cite{nagabandi2018neural}, the World Models algorithm \cite{ha2018recurrent}, the I2A algorithm \cite{racaniere2017imagination}, and etc. 

The model-free RL does not predict the environment parameters but directly seeks the optimal policy using the samples. The idea is to look constantly for the policy that produces higher rewards. An example is the $Q$-learning algorithm where the agent chooses, with a high probability, the action corresponding to the highest $Q$ value. The $Q$-algorithm converges to an optimal $Q$-value function under proper conditions, which produces an optimal policy. The difference between model-based and model-free RL is whether the agent needs to construct/estimate the model of the environment (e.g., the transition kernel and the cost function). In Section \ref{subsec:attacksReward}, we will discuss the vulnerabilities of model-based RL and model-free RL under adversarial attacks on the cost samples.

\subsubsection{Value-Based and Policy-Based}
The model-free RL has two main categories for optimizing the policy: value-based methods and policy-based methods. The value-based methods usually employ a value-action function $Q:\mathcal{S}\times\mathcal{A}\rightarrow \mathbb{R}+$, which is also referred to as the $Q$ function. The most population algorithm of the value-based method is the $Q$-learning algorithm, whose goal is to find the optimal $Q$-values that satisfy the Bellman equation \cite{bertsekas1996neuro} 
}}
$$
Q(s,a) = c(i,a) + \beta \sum_{s'} p(s,s',a) \min_{a'}Q(s',a'),\ \ \ \textrm{for } i\in\mathcal{S}, a\in\mathcal{A},
$$ 
where $p(s,s',a)$ is the probability that the state of the cyber system at the next step is $s'$ given the current state $s$ and the current action $a$. Without the knowledge of the transition probability $p(\cdot, \cdot, \cdot)$ and the cost function $c(\cdot,\cdot)$, the RL agent can update its $Q$-values by interacting with the environment:
\begin{equation}\label{eq:QAlgo}
Q_{t+1}(s_t,a_t)  =  Q_t(s_t,a_t) + \alpha_t \cdot \left[ \beta \min_{a'}Q_n(s_{t+1},a') + c_t - Q_t(s_t,a_t) \right], 
\end{equation}
where the sequences of states $\{s_t\}_{t\in \mathbb{Z}}$, costs $\{c_t\}_{t\in\mathbb{Z}}$ are from the environment and the sequence of actions $\{a_t\}_{t\in\mathbb{Z}}$ are chosen by the agent.

{{The policy-based RL focuses on the policy directly. The policy-based method updates the policy iteratively to minimize the accumulative costs over a period of time without explicitly constructing a value function. The adoption of policy-based RL allows the agent to parameterize the policy, which often results in better convergence and is suitable for continuous or high dimensional action space. Frequently used policy-based algorithms include the Policy Gradient (PG) algorithm proposed by Sutton et al. \cite{sutton2000policy}, the Trust Region Policy Optimization (TRPO) algorithm \cite{schulman2015trust}, etc.}

Combining the value-based and the policy-based methods gives rise to the actor-critic class of algorithms. The actor-critic method leverages the value-based methods to build a value function to improve sample efficiency and relies on the policy-based methods to update the policy parameterization. Common actor-critic RL algorithms include the Actor-Critic (AC) algorithm \cite{sutton2018reinforcement} and its variants: the Asynchronous Advantage Actor-Critic (A3C) algorithm \cite{mnih2016asynchronous}, the Deterministic Policy Gradient (DPG) algorithm \cite{silver2014deterministic}, etc.

\subsubsection{Function Approximation}
With the curse of modeling being tamed, the next is to discipline the curse of dimensionality. The curse of dimensionality appears when the state space or the action space is continuous or prohibitively large, which is the situation we often encounter in cyber systems. Function approximation is a technique that addresses the curse of dimensionality. A function approximator is selected to approximate the value function or the policy function so that the learning algorithms only have to deal with a few parameters. Common function approximators are linear function approximator \cite{bertsekas1996neuro,sutton2018reinforcement,huang2020manipulating}, multiscale approximator \cite{li2019convergence}, neural networks \cite{nguyen2019deep,mnih2016asynchronous,franccois2018introduction,sutton2018reinforcement}, etc. Recent development in Deep Neural Networks (DNNs) in the last decade enables the application of Deep Reinforcement Learning (DRL). In DRL, the function approximator is a neural network with multiple layers between the input and output layers. A wide variety of DNNs, such as convolutional neural networks, recurrent neural networks \cite{franccois2018introduction} and transformers \cite{vaswani2017attention}, has been developed to handle different application scenarios. One can refer to Section 7 of \cite{franccois2018introduction} and Section 9 of \cite{sutton2018reinforcement} for a detailed discussion on the selection of function approximators.

\subsubsection{Other RL Settings}
In the training phase, the RL agent chooses a policy that interacts with the environment to generate more samples and constantly improve the policy using the generated samples. The RL agent accumulates knowledge about the cyber system. The agent has to make a trade-off between learning more about the systems or pursuing what seems to be the most promising defense strategy with the samples gathered so far. This is called the exploration-exploitation trade-off in choosing how to sample from the environment (or the cyber systems in our case). Commonly used sampling strategies include the greed strategy (i.e., choose the action currently unknown as the best), the $\epsilon$-greedy exploration (i.e., applies the greedy action with probability $1-\epsilon$ and otherwise selects an action uniformly at random), or the Thompson sampling \cite{russo2018tutorial}.

In many RL applications, the agent is equipped with a replay memory that stores the past experience/samples of the agent. The storage of past experience allows the agent to reuse the samples later and process samples in batches to achieve a reasonably stable data distribution \cite{franccois2018introduction}. With replay memory, the agent can choose either on-policy or off-policy methods. On-policy methods evaluate and improve the policy using the samples generated by the same policy. In off-policy methods, learning is done using samples in replay memory that are not necessarily generated under the current policy but different policies. The off-policy methods allow the defender to leverage data from historical security events to learn a defense policy. A typical example of the on-policy method is SARSA \cite{sutton2018reinforcement} where the agent executes an action from a $\epsilon$-greedy policy and uses the sample generated by the action to update the current policy:
\begin{equation}\label{eq:Sarsa}
Q_{t+1}(s_t,a_t)  =  Q_t(s_t,a_t) + \alpha_t \cdot \left[ \beta Q_n(s_{t+1},a_{t+1}) + c_t - Q_t(s_t,a_t) \right].
\end{equation}
$Q$-learning algorithm in (\ref{eq:QAlgo})) is a frequently used off-policy method. Even though it adopts the $\epsilon$-greedy policy to generate actions, the update of $Q$-values follows the greedy policy as is indicated by the min operator in (\ref{eq:QAlgo}). With basic knowledge of RL methods, in the following section, we discuss the literature about how RL can be used to tackle different attacks in cyber systems. 
}

\subsection{Review of RL Against Cyber Attacks}\label{subsec: rev_rl_algs} 

The rapid development of Information and Communication Technologies (ICTs) has fueled significant growth in Internet of Things (IoT) usage and Cyber-Physical Systems (CPS) deployment.  CPS involves embedded computers and networks used to monitor and control the physical processes, with feedback loops where physical processes affect computations and vice versa. The Internet of Things (IoT) refers to a network comprised of physical objects capable of gathering and sharing electronic information, with physical objects receiving orders from the network and the network receiving information from the physical objects. Both IoT and CPS involve computation units, networking units, and physical entities that form a feedback cycle with human or machine decisions in the loop. Due to the wide adoption of IoT and CPS in many domains such as manufacturing, power, agriculture, battlefield, homes, etc., security threats faced by IoT and CPS systems need to be well addressed. 

First, in a CPS, traditional physical controllers have been replaced by micro-computers and processors with embedded operating systems \cite{cardenas2008research}. Micro-computers provide advantages to the CPS, such as flexible configuration with an accessible human-machine interface and digital communication abilities that enable remote control and access. However, the embedded software opens up doors for software attacks such as code injection attacks, cross-site scripting attacks when using a web server for system configuration, and malware attacks. Second, CPS is networked, meaning that the essential components in the physical systems are often connected to corporate networks and the Internet. Security issues may be raised because complete isolation is difficult to achieve. Connectivity can occur in unexpected ways. For example, Stuxnet, a malicious computer worm accused of stymieing Iran’s nuclear-fuel enrichment program, was introduced to the target environment via an infected USB flash drive \cite{langner2011stuxnet}. The security challenges of CPS will become more severe as the scale and scope of the Internet and other communication techniques grow.

Similarly, IoT is also facing security challenges amid the increasing adoption of IoT technology. IoT employs advanced computing and communication technology such as Zigbee, Bluetooth, radio-frequency identifications (RFIDs), and cloud computing. The employment of these technologies provide opportunities for the attacker to launch cyber attacks, such as DoS attack, eavesdropping, data injection attacks, etc. Beyond that, unlike CPS, IoT usually encompasses small physical units with lower power that are mobile and geographically distributed. Hence, IoT is more susceptible to physical attacks. Due to the dynamic and feedback nature of CPS and IoT, RL has been applied to address the security and privacy challenges faced by these systems. In this section, we provide a brief review of the applications of RL for the security of CPS and IoT. The review follows a taxonomy based on the attack types and different RL algorithms.

\subsubsection{RL Aganist DoS attacks}

DoS is the most common attack among all the attacks targeting networked systems such as IoT and CPS  \cite{alanazi2015resilience,cardenas2008secure}. A denial-of-service (DoS) attack is a tactic used by attackers for overloading a machine or network to make it unavailable. DoS attacks on networked systems bring serious threats to people and induce direct or indirect financial losses. Since many CPS relies heavily on real-time communication, the sudden unavailability of information may render serious issues. For example, suppose a critical chemical reaction process is unstable in open-loop control. In this case, a continuous DoS attack on the actuator may introduce irreversible damage to the system and the beings around it. Liu et al. \cite{liu2020event} have studied the resilient control problem for CPS under DoS attacks. The DoS attacks target the measurement and control channels to jeopardize the closed-loop system's functionality. The authors designed a resilient control strategy that can achieve a certain degree of stability and performance for the CPS even when under DoS attacks. To compute such a resilient control strategy, the authors have employed an on-policy RL method called SARSA to calculate the strategy. Dai et al. \cite{dai2020distributed} have studied the remote state estimation of CPS with multiple sensors. The sensors transmit the message through several communication channels subject to possible DoS attacks. The authors leverage $Q$-learning to compute the communication strategy, i.e., to decide which communication channel to use.

Due to its distributed nature with limited energy, IoT is often targeted by Distributed DoS (DDoS) attacks. Distributed DoS often results in malfunction in an enormous number of IoT devices within a short time window. Xu et al. \cite{xu2007defending} have proposed a DDoS detection scheme based on Hidden Markov Models and cooperative RL. The authors leverage DQN to compute optimized strategies of information exchange among the distributed multiple detectors to ensure a decent detection accuracy without much load on information communications among the detectors. Later in 2013, Malialis et al. \cite{malialis2013large} have proposed a protective mechanism to protect IoTs from DDoS attacks by router throttling. A multi-agent RL framework is applied where the agents representing the routers learn to throttle traffic toward a victim server. However, this method will suffer scalability issues as the scale of IoT increases. To tackle this problem, the same authors proposed a coordinated design on top of the previous RL-based router throttling method \cite{malialis2015distributed}. More recently, Liu et al. have proposed a DQN based framework to mitigate the effect of the DDoS attack, which can increase or slow down the attack traffic flow \cite{liu2018deep}. Most papers that leverage RL to mitigate the effect of DoS attacks are for computational purposes when the behavior model of the attacker is unknown to the agent.  

\subsubsection{RL Aganist Spoofing Attacks}
A spoofing attack happens when a malicious party impersonates another device/user on a network to launch other attacks, steal data, spread malware, or bypass access control \cite{pradhan2016gps}. There are many things the attackers can forge to make their attacks pan out: An IP address, a phone number, GPS location, etc. Out of all the nefarious scenarios, the following few become increasingly impactful for cyber security nowadays. ARP (Address Resolution Protocol) spoofing is a spoofing attack in which a malicious attacker sends falsified ARP messages over a local area network. This {manipulation} can make all traffic redirected to the attacker's device before it can reach its legitimate destination. Another common type of spoofing attack is the IP spoofing attack, in which the attacker impersonates the IP address and obfuscates the actual online identity of the packet sender. The penetration using spoofing attacks gives wings to the attacker to launch further attacks such as man-in-the-middle or DoS attacks. 

Xiao et al. have formed a zero-sum authentication game to capture the interplay between the legal receiver and spoofers \cite{xiao2015spoofing,xiao2016phy}. The authors leverage $Q$-learning algorithms to compute the spoofing detection strategy. The receiver uses this strategy to choose the test threshold in the hypothesis test of the spoofing detection to maximize its expected utility based on Bayesian risk to detect the spoofer. Beyond the passive defensive method such as detection, Elnaggar et al. \cite{elnaggar2018irl} have proposed an inverse RL approach that leverages historical measurement and control data to forecast the objective of spoofers. The authors studied an autonomous vehicle equipped with multiple sensors and tasked to perform go-to-goal navigation. The spoofer aims to hijack the vehicle to an adversary-desired location while staying stealthy. The agent can prevent the vehicle from reaching the adversary-desired location by leveraging the IRL method.

\subsubsection{RL Aganist False Data Injection Attacks}

False data injection (FDI) attacks affect the data integrity of packets by modifying their payloads, which are in general more difficult to detect but have less been investigated \cite{mo2010false}.  Li et al. \cite{li2019deep} proposed a DQN-based RL method, on behalf of the attacker, to jeopardize the data analysis results by optimally injecting fake data conflicting with ground truth in crowdsensing systems. Similarly, In \cite{chen2018evaluation}, Chen et al. come up with a novel strategy of false data injection (FDI) attacks on behalf of the attacker, aiming to distort the regular operation of a power system by automatic voltage controls. The problem of finding an optimal attack strategy is formed as a partially observable MDP, and a Q learning algorithm is proposed to enable online learning and attacking.

Instead of designing optimal FDI attacking strategies for the attacker, Kurt et al. \cite{kurt2018online} designed a defense strategy against a series of cyber attacks in smart grids. The authors proposed a model-free RL algorithm to detect cyber attacks, including  FDI attacks, on the fly without knowing the underlying attack model. Numerical results show that the proposed detection system can detect FDI attacks with a precision of $0.9977$ and a recall of $1$. This section has introduced different types of attacks, how they can cause damage to the systems, and how RL can mitigate the damage. The following section focuses on the three major classes of vulnerabilities in cyber systems and discusses how RL can help address each class of vulnerabilities.




% ...

 

 
 
 

\section{Resilient Cyber Defense Against Vulnerabilities}
\label{sec:reviewRLCR}


As discussed in Section \ref{sec:intro}, the preparation and prevention mechanisms for perfect security are too costly to implement. Response and recovery mechanisms are necessary to mitigate successful attacks and harden the system security.  RL is an efficient tool to achieve this goal through online learning, correction, and adaptation. 
In this section, we provide a literature overview on cyber resilience and focus on methodologies that rely on feedback mechanisms, which include RL and control-theoretic approaches. We refer readers to several recent reviews on game theory for cybersecurity \cite{manshaei2013game} and cyber deception \cite{pawlick2019game}, and Deep Reinforcement Learning (DRL) for cybersecurity \cite{nguyen2019deep}. The design of CRMs depends on the type of vulnerabilities that the cyber system aims to mitigate. We group the common vulnerabilities into three categories, namely, (1) the posture-related vulnerabilities, (2) the information-related vulnerabilities, and (3) the human-related vulnerabilities. 

The posture-related vulnerabilities refer to a defender's naturally disadvantageous security posture in comparison to the attacker's; e.g., the defender has to be constantly vigilant and legitimately defend the entire attacker surface, while the attacker only needs to succeed once at a single location. 
Due to the disadvantage in security posture, the defender with limited resources cannot afford to prepare for all possible attacks. 
The information-related vulnerabilities refer to the defender's information disadvantage, especially when facing deceptive and stealthy attacks. The defender cannot make a meticulous plan to protect his assets if he does not have the capability of mapping out the attack paths or predicting the targets. {The human-related vulnerabilities are the results of human misbehavior and cognition limitation. 
The vulnerabilities of all human groups in the cyber system can expose the system to cyber threats and undermine cyber resilience. 
Human users and insiders can unintentionally fall victim to phishing attacks or intentionally break security rules for their convenience. 
Human operators and network administrators in charge of real-time monitoring and inspections of alerts and system status can suffer from alert fatigue.}  
Human vulnerabilities are often exploited by attackers as the first step who aim to penetrate a computing system. 
Preparation and prevention mechanisms, e.g., security training and regulations, are not sufficient to protect them from attacks. 

{We briefly review the existing technologies to mitigate the three types of  vulnerabilities in Section \ref{sec:posVul}, \ref{sec:InfoVul}, and \ref{sec:HumVul}, respectively. 
In particular, we emphasize the RL-based CRMs and introduce three cyber resilience designs, i.e., MTD in Section \ref{sec: MTD}, honeypots in Section \ref{sec:honeypot}, and human-assistive technologies in Section \ref{sec:human}, as the representative examples of the RL solutions to these three types of vulnerabilities, respectively. 
}

\subsection{Posture-Related Vulnerability}\label{sec:posVul}

The technologies to mitigate posture-related vulnerabilities have been extensively studied in the literature. 
Moving Target Defense (MTD) is one of the modern technologies to neutralize attacker's position advantage by creating reconnaissance difficulties and uncertainties for attackers. 
There is a surge of recent literature on using RL to choose an adaptive configuration strategy to maximize the impact of MTD with particular focuses on the dynamic environment \cite{gao2021reinforcement}, reduced resource consumption \cite{chai2020dq}, usability \cite{zhu2013game}, partially observable environment \cite{yoon2021desolater,sengupta2020multi}, and multiagent scenarios that contains both the characteristics of the system and the adversary’s observed activities \cite{eghtesad2020adversarial,sengupta2020multi}.

Other techniques have been proposed to protect disadvantageous cyber systems from falling victim to resourceful and determined attacks such as Distributed Denial-of-Service (DDoS) or powerful jamming attacks. 
For example, \cite{liu2018deep} mitigates DDoS Flooding in Software-Defined Networks; \cite{feng2020application} introduces a multi-objective reward function to guide an RL
agent to learn the most suitable action in mitigating application-layer DDoS attacks; \cite{li2021ddos} proposes a feature adaption RL approach based on the space-time flow regularities in IoV for DDoS mitigation. 
Among the works of RL for DDoS attack, there has been recent attention on large-scale solution \cite{malialis2013large,xu2007defending} via cooperative RL, low-rate attacks in the edge environment \cite{liu2020cpss}, and sparse constraint in cloud computing \cite{zhu2021power}. 
Besides DDoS attacks, RL has a wide application in mobile edge caching \cite{xiao2018security}, mobile offloading for cloud-based malware detection \cite{wan2017reinforcement}, and host-based and network-based intrusion detection \cite{servin2005multi,LOPEZMARTIN2020112963}.

\subsection{Information-Related Vulnerability}
\label{sec:InfoVul}
{Deception is a ubiquitous phenomenon and has been used as a proactive defense method.}
Honeypots and the related {cyber deception} technologies, e.g., honeyfiles, honeynet, honeybot, etc., have been widely used to mitigate stealthy and deceptive attacks. 
In particular, RL has been widely used to develop self-adaptive honeypots in normal conditions  \cite{pauna2018qrassh,pauna2019rewards,wagener2011adaptive} and resource-constrained environment \cite{venkatesan2017detecting}. 
As attackers become more sophisticated, they manage to fingerprint and evade honeypots \cite{krawetz2004anti}. This reduces the size of a captured dataset and the chance to extract threat intelligence from it. 
Thus, there is an urgent need for stealthy honeypots \cite{huang2020farsighted} that can counter fingerprinting. 
In \cite{dowling2018using}, the authors use RL to conceal the honeypot functionality. 
Besides honeypot fingerprinting, RL is also used to extract the most threat intelligence \cite{dowling2018improving,suratkar2021adaptive,huang2019adaptive}. 

Besides honeypots, there are emerging technologies to detect and respond against deceptive attacks.
In \cite{RN649,huang2019adaptive2,RN660}, the authors model the interactions between stealthy attackers and the defenders as a dynamic Bayesian game of discrete or continuous type where multi-stage response strategies are designed based on belief updates. The convergence of continuous-type Bayesian game under a sequence of finer discretization schemes is discussed in \cite{huang2021convergence}. 
In \cite{huang2020game}, the authors design the information revealed to the users and attackers to elicit behaviors in favor of the defender.
In \cite{wang2020intelligent} and \cite{bhattacharya2020automated}, RL is used to optimally deploy the deception resources and emulate adversaries, respectively. 
Many works have attempted to address various spoofing attacks using RL. 
In \cite{xiao2016phy}, the authors propose RL-based spoofing detection schemes where the interactions between a legitimate receiver and spoofers are formulated as a zero-sum authentication game. 
In \cite{cai2020drl}, the authors model the behavior of exploring face-spoofing-related
information from image sub-patches by leveraging deep RL. 
Developing counter-deceptive and defensive deceptive technologies is in its infancy. RL is a promising tool to make the design {quantitative, automatic, and adaptive}.

\subsection{Human-Related Vulnerability}
\label{sec:HumVul}
{
We classify human vulnerabilities into acquired and innate vulnerabilities, depending on whether they can be mitigated through short-term security training and awareness programs.} 
Social engineering \cite{salahdine2019social} is a common attack vector that targets acquired human vulnerabilities such as fear to express anger, lack of assertiveness to say no, and the desire to please others. 
Threat actors use psychological manipulation techniques to mislead people to break normal security procedures or divulge confidential information. 
The authors in \cite{aind2020q} have used RL to detect cyberbullying automatically, and \cite{yang2018use,gonzalez2006framework} have proposed a feedback learning framework, e.g., RL, to fight against social engineering attacks. 
As a representative form of social engineering,  phishing attacks use email or malicious websites to serve malware or steal credentials by masquerading as a legitimate entity. 
Non-technical anti-phishing solutions include security training and education programs, while technical solutions include blacklisting, whitelisting, and feature-based detection. 
To handling zero-day phishing attacks, (deep) RL has been used both to detect phishing emails \cite{smadi2018detection}, phishing websites \cite{chatterjee2019detecting}, spear phishing \cite{evans2021raider}, and social bots \cite{lingam2019deep} in Online Social Networks (OSN). 
{Attackers can also exploit human innate vulnerabilities induced by bounded attention and rationality. 
The authors in \cite{RN661,huang2021radams} have identified a new type of attacks called Informational Denial-of-Service (IDoS) attacks that can exacerbate the human operators' cognition overload by generating a large number of feints and hiding real attacks among them.}

Most of the existing works take humans as an independent component in the cyber system and aim to compensate indirectly for the human vulnerability through additional mechanisms. 
{An alternative way is} to directly affect the human component and consider an integrated \textit{human-cyber system}. 
Due to the unpredictability and modeling challenges of human behaviors, RL and feedback control serves as the tool to determine how to affect human incentives and perceptions effectively and efficiently. 
In \cite{casey2015compliance}, the penalty and reward are changed adaptively through a feedback system to improve compliance of human employees and mitigate insider threats. 
In \cite{huang2021inadvert}, RL is used to develop the optimal visual aids to enhance users' attention and help them identify phishing attacks. 
{In \cite{huang2021radams}, the authors use RL to determine resilient and adaptive strategies for alert and attention management.}  
Developing security-assistive technologies that directly affect humans is in its infancy, and it is a promising direction to explore further.


\section{Cyber-Resilient Mechanism Designs and Applications}
\label{sec:CRMechanismAndApplications}

The system vulnerabilities and attacker's information and resource advantages have made the response and recovery mechanisms a necessity. 
To achieve cyber resilience, the defender needs to make the response and recovery mechanisms \textit{configurable, adaptive, and autonomous}. RL and feedback control provide a rich set of tools to achieve these three features. 

In this section, we introduce the following three cyber resilience designs, i.e., {MTD strategies for security-usability trade-off \cite{zhu2013game}} in Section \ref{sec: MTD}, {honeypot strategies for attack engagement \cite{huang2019adaptive}} in Section \ref{sec:honeypot}, and {human-assistive alert management strategies \cite{RN661,huang2021radams}} in Section \ref{sec:human}, as the representative examples of RL solutions to three vulnerabilities in Sections \ref{sec:posVul}, \ref{sec:InfoVul}, and \ref{sec:HumVul}, respectively. 
For each application, we first introduce the necessary background, motivation, and challenges of the defense technologies. 
Then, we introduce the feedback and learning models to capture the online behaviors of the players, e.g., the defender, the attackers, and the users. 
Finally, we illustrate how RL can enable cyber resilience. 

{
The CRMs in all three works guarantee asymptotic convergence, but the learning speed varies based on the complexity of the objective strategy and the learning environment. 
Fast convergence can reduce the response time in Fig. \ref{crm} and lead to a more resilient solution to enhance cyber security performance. 
Many works in the RL literature, including approximations \cite{barreto2020fast}, transfer learning \cite{zhuang2020comprehensive}, and meta learning \cite{lemke2015metalearning}, have improved the learning speed and reduced sample complexity. 
However, few works have focused on applying the above success to design CRM. 
We envision seeing more works that apply those RL algorithms to the cyber domain and thus enable efficient learning and timely responses. 
Besides developing general efficient RL algorithms, we can also tailor and improve the existing RL solutions to the characteristics of cyber systems. 
For example, it is sufficient to consider epsilon-optimality when the cyber system is not stationary; e.g., its state changes dynamically and stochastically based on the behaviors of users, defenders, and attacks. 
}

\subsection{Adaptive MTD Strategies for Security-Usability trade-off} 
\label{sec: MTD}

Many legacy Information Technology (IT) systems operate in a relatively static configuration. 
As many configuration parameters, e.g., IP addresses, remain the same over long periods, advanced attackers can reconnoiter and learn the configuration to launch targeted attacks. 
 MTD is a countermeasure that dynamically configures the system settings and shifts the attack surface to increase uncertainties and learning costs for the attackers. 

Many networked IT systems nowadays consist of hierarchical layers and adopt the Defense-in-Depth (DiD) approach where a series of defensive mechanisms reside on each of these layers. 
Let $\mathcal{N}:=\{1,2,\cdots,N\}$ denote the set of $N$ layers in a system and $\mathcal{V}_l:=\{v_{l,1},v_{l,2},\cdots,v_{l,n_l}\}$ be the set of $n_l$ system vulnerabilities that an attacker can exploit to compromise the system at layer $l\in \mathcal{N}$. 
Assuming that perfect security is unattainable and attacks can penetrate the layered defense of different depths, MTD can be applied for all layers to interact and slow down the attacker's rate of penetration. 
At each layer $l\in \mathcal{N}$, the defender can choose to change the system configuration from a finite set of $m_l$ feasible configurations $\mathcal{C}_l:=\{c_{l,1},c_{l,2},\cdots,c_{l,m_l}\}$. 
Different configurations result in different subsets of vulnerabilities among $\mathcal{V}_l$, which are characterized by the configuration-vulnerability map $\pi_l:\mathcal{C}_l \rightarrow 2^{\mathcal{V}_l}$. 
Under the $h$-th configuration $c_{l,h}\in \mathcal{C}_l$, the outcome of the configuration-vulnerability mapping, i.e., $\pi_l(c_{l,h})$, represents the attack surface  at layer $l\in \mathcal{N}$.

After the attacker has reached the layer  $l\in \mathcal{N}$, the attacker can launch an {attack} $a_{l,k}$ from a finite set $\mathcal{A}_l:=\{a_{l,1},a_{l,2},\cdots,a_{l,n_l}\}$. 
Let $\gamma_l: \mathcal{V}_l \rightarrow \mathcal{A}_l$ be the vulnerability-attack map that associates vulnerability $v_{l,j}\in  \mathcal{V}_l$ with attack $a_{l,k}\in \mathcal{A}_l$. 
Without loss of generality, there is a one-to-one correspondence between $\mathcal{A}_l$ and $\mathcal{V}_l$, then the corresponding inverse map is denoted as $\gamma_l^{-1}: \mathcal{A}_l \rightarrow \mathcal{V}_l$. 
Attack action $a_{l,k}\in \mathcal{A}_l$ incurs a bounded cost $D_{hk}\in \mathbb{R}_{+}$ when  the current attack surface $ \pi_l(c_{l,h})$ under configuration $c_{l,h}\in \mathcal{C}_l$ contains the vulnerability $v_{l,j}=\gamma_l^{-1}(a_{l,k})$. 
Otherwise, the attacker fails to exploit the existing vulnerabilities under the configuration and incurs zero damage. 
Thus, the damage caused by the attacker at layer $l\in \mathcal{N}$, denoted by $r_l: \mathcal{A}_l \times \mathcal{C}_l \rightarrow \mathbb{R}_+$, takes the following form:
  \begin{equation}
    r_l(a_{l,k},c_{l,h})=
    \begin{cases}
      D_{hk}, &  \gamma_l^{-1}(a_{l,k})\in \pi_l(c_{l,h}) \\
      0, & \text{otherwise}
    \end{cases}.  
  \end{equation}

\begin{table}[H]
    {
    \caption{A table of parameters for Section \ref{sec: MTD}}
    \begin{tabularx}{\columnwidth}{X l} 
    \toprule
      {\underline{Indices \& Sets}: } \\
      $N, \mathcal{N}$ & Number and set of layers \\
      $n_l,\mathcal{V}_l$ & Number and set of vulnerabilities at layer $l$: $\{v_{l,1}, v_{l,2},\cdots, v_{l,n_l}\}$\\
      $\mathcal{C}_l$ & Set of system configurations at layer $l$: $\{c_{l,1},c_{l,2},\cdots,c_{l,m_l}\}$\\
      $\mathcal{A}_l$ & Set of attacks available at layer $l$: $\{a_{l,1},a_{l,2},\cdots,a_{l,n_l}\}$\\
      
    {\underline{Decision \& Strategy}: } \\
      $a_{l,k}$ & Attack targeting at vulnerability $k$ at layer $l$\\
      $\Delta \mathcal{C}_l, \Delta \mathcal{A}_l$ & Distributions of set $\mathcal{C}_l$ and $\mathcal{A}_l$ \\
      $\mathbf{f}_l =\{f_{l,1},f_{l,2},\cdots,f_{l,m_l}\}\in \Delta \mathcal{C}_l$ & Defender's randomized strategy over $\mathcal{C}_l$\\
      $\mathbf{g}_l =\{g_{l,1},g_{l,2},\cdots, g_{l,n_l}\} \in \Delta \mathcal{A}_l$ & Attacker's randomized attacking strategy over $\mathcal{A}_l$\\
      $\mathbbm{c}_{l,t}(\mathbbm{a}_{l,t})$ & Action chosen by defender (attacker) at time $t$\\

      {\underline{Other Variables}: } \\
      $\pi_l:\mathcal{C}_l\rightarrow 2^{\mathcal{V}_l}$ & Configuration-vulnerability mapping\\
       $\gamma_l:\mathcal{V}_l \rightarrow \mathcal{A}_l$ & Vulnerability-attack map\\
      $\gamma_l^{-1}:\mathcal{A}_l \rightarrow \mathcal{V}_l$ & Inverse vulnerability-attack map \\
      $D_{hk}$ & Cost induced by the attack indexed by $k$ under $h$-th configuration \\
      $r_l:\mathcal{A}_l \times \mathcal{C}_l \rightarrow \mathbb{R}_+$ & Damage caused by the attacker at layer $l$\\
      $\hat{r}_{l,t}^S (\hat{r}_{l,t}^A)$ & Defender's (attacker's) estimate of the average risk \\
      $\mu_t^S (\mu_t^A)$ & Payoff learning rates for the defender (attacker)\\
      $\mathbbm{r}_l(\mathbf{f}_l,\mathbf{g}_l)$ & Defender's expected cost under strategies $\mathbf{f}_l$ and $\mathbf{g}_l$\\
      $R_{l,t}^S$ & Reconfigure cost of the defender\\
      $\epsilon_{l,t}^S$ & Parameter that balances the security and the usability\\
      $W_{l,t}^S(W_{l,t}^A)$ & Optimal value of the defender's (attacker's) optimization problem at time $t$\\
      $\lambda_{l,t}^S(\lambda_{l,t}^A)$ & Learning rate of the defender (attacker)\\
      \bottomrule
     \end{tabularx}
     }
    \end{table}

\begin{figure}[ht]
    \centering
    \begin{subfigure}[]{0.45\textwidth}
        \centering
        \includegraphics[width=0.6 \textwidth]{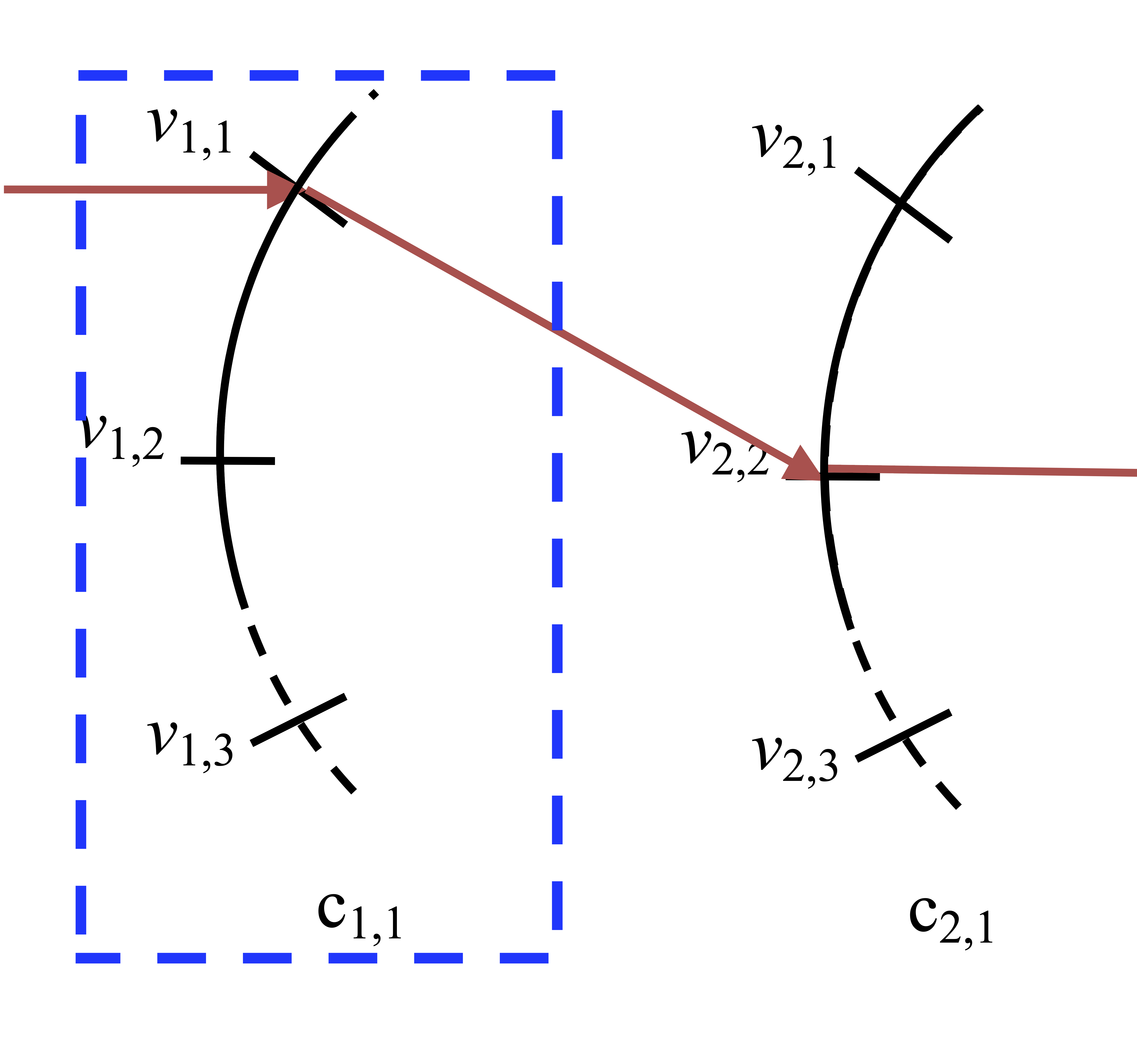}
        \caption{Attack surface $\pi_1(c_{1,1})=\{v_{1,1},v_{1,2}\}$ under configuration $c_{1,1}$. }
        \label{fig:config1}
    \end{subfigure}%
    \hfill 
    \begin{subfigure}[]{0.45\textwidth}
        \centering
        \includegraphics[width=0.6 \textwidth]{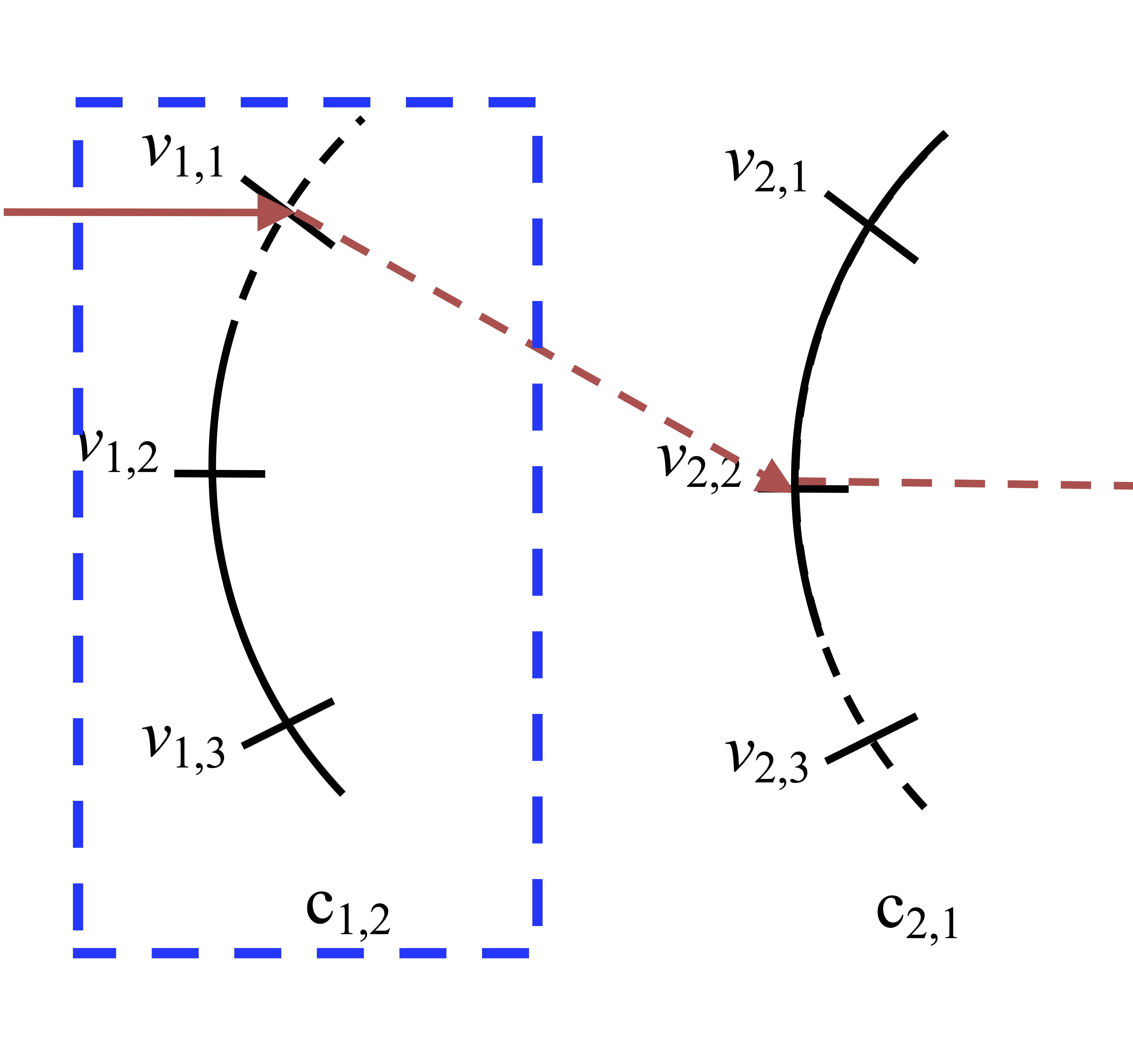}
        \caption{Attack surface $\pi_1(c_{1,2})=\{v_{1,2},v_{1,3}\}$ under configuration $c_{1,2}$. }
         \label{fig:config2}
    \end{subfigure}
    \caption{
    Given a static configuration $c_{1,1}\in \mathcal{C}_1$, an attacker can succeed in reaching the resources at deeper layers by forming an attack path from  $v_{1,1}$ to $ v_{2,2}$. A change of configuration to $c_{1,2}\in \mathcal{C}_1$ can prevent the attacker from exploiting the vulnerabilities at the first layer. 
 }
  \label{fig:config12}
\end{figure}

The attacker's goal is to penetrate and compromise the system, while the defender aims to minimize the damage or risk. 
Since vulnerabilities are inevitable in modern IT systems, the defender adopts MTD to randomize between configurations and make it difficult for the attacker to learn and exploit the vulnerabilities at each layer.  
Fig. \ref{fig:config12} provides a paradigmatic example to illustrate the benefit of MTD.  
At the first layer highlighted by the blue box, there are three vulnerabilities. 
Configuration $c_{1,1}$  in Fig. \ref{fig:config1} has an attack surface $\pi_1(c_{1,1})=\{v_{1,1},v_{1,2}\}$ while configuration $c_{1,2}$ in Fig. \ref{fig:config2} has an attack surface $\pi_1(c_{1,2})=\{v_{1,2},v_{1,3}\}$. 
For each attack surface, the existing and non-existing vulnerabilities are denoted by the solid and dashed arcs, respectively. 
Then, if the attacker takes action $a_{1,1}\in \mathcal{A}_1$ that exploits vulnerability $v_{1,1}$ but the defender changes the configuration from $c_{1,1}$ to $c_{1,2}$, then the attack is thwarted at the first layer. 

{The authors in \cite{zhu2013game}} model the conflict between a multi-stage attacker and a defender adopting MTD as a zero-sum game. 
At each layer $l\in \mathcal{N}$, the defender's randomized strategy over the configuration set $\mathcal{C}_l$ is denoted as $\mathbf{f}_l \coloneqq\{f_{l,1},f_{l,2},\cdots,f_{l,m_l}\}\in \Delta \mathcal{C}_l$ where $ \Delta \mathcal{C}_l$ is the distribution of set $\mathcal{C}_l$. 
The attacker can also choose a randomized strategy  $\mathbf{g}_l:=\{g_{l,1},g_{l,2},\cdots,g_{l,n_l}\}\in \Delta \mathcal{A}_l$ over the set of feasible attacks at layer $l\in \mathcal{N}$ to increase his chance of a successful vulnerability compromise. 
In particular,  $f_{l,h}\in [0,1]$ and  $g_{l,k}\in [0,1]$ represent the probabilities of the defender taking configuration $c_{l,h}$ and the attacker taking action $a_{l,k}$, respectively, at layer $l\in \mathcal{N}$. 
Under the mixed-strategy pair $(\mathbf{f}_l\in  \Delta \mathcal{C}_l,\mathbf{g}_l\in  \Delta \mathcal{A}_l)$, the defender's expected cost  $\mathbbm{r}_l$ is given by
\begin{equation}
\mathbbm{r}_l(\mathbf{f}_l,\mathbf{g}_l) :=\mathbb{E}_{\mathbf{f}_l,\mathbf{g}_l}r_l =\sum_{h=1}^{m_l}\sum_{k=1}^{n_l} f_{l,h}g_{l,k}r_l(a_{l,k},c_{l,h}). 
\end{equation}

The defender's optimal randomized strategy $\mathbf{f}_l^*\in  \Delta \mathcal{C}_l$ against the worst-case attacks can be obtained from the mixed strategy saddle-point equilibrium (SPE) $(\mathbf{f}_l^*\in  \Delta \mathcal{C}_l,\mathbf{g}_l^*\in  \Delta \mathcal{A}_l)$ of the zero-sum game where the game value $\mathbbm{r}(\mathbf{f}_l^*,\mathbf{g}_l^*)$ is unique, i.e., 
\begin{equation}
\label{eq:SPE}
\mathbbm{r}_l(\mathbf{f}_l^*,\mathbf{g}_l)\leq \mathbbm{r}_l(\mathbf{f}_l^*,\mathbf{g}_l^*) \leq \mathbbm{r}_l(\mathbf{f}_l,\mathbf{g}_l^*), \forall \mathbf{f}_l \in  \Delta \mathcal{C}_l, \mathbf{g}_l \in  \Delta \mathcal{A}_l, 
\end{equation}

\subsubsection{Optimal Configuration Policy via Reinforcement Learning}
\begin{figure}[ht]
\centering
\includegraphics[width=0.6 \textwidth]{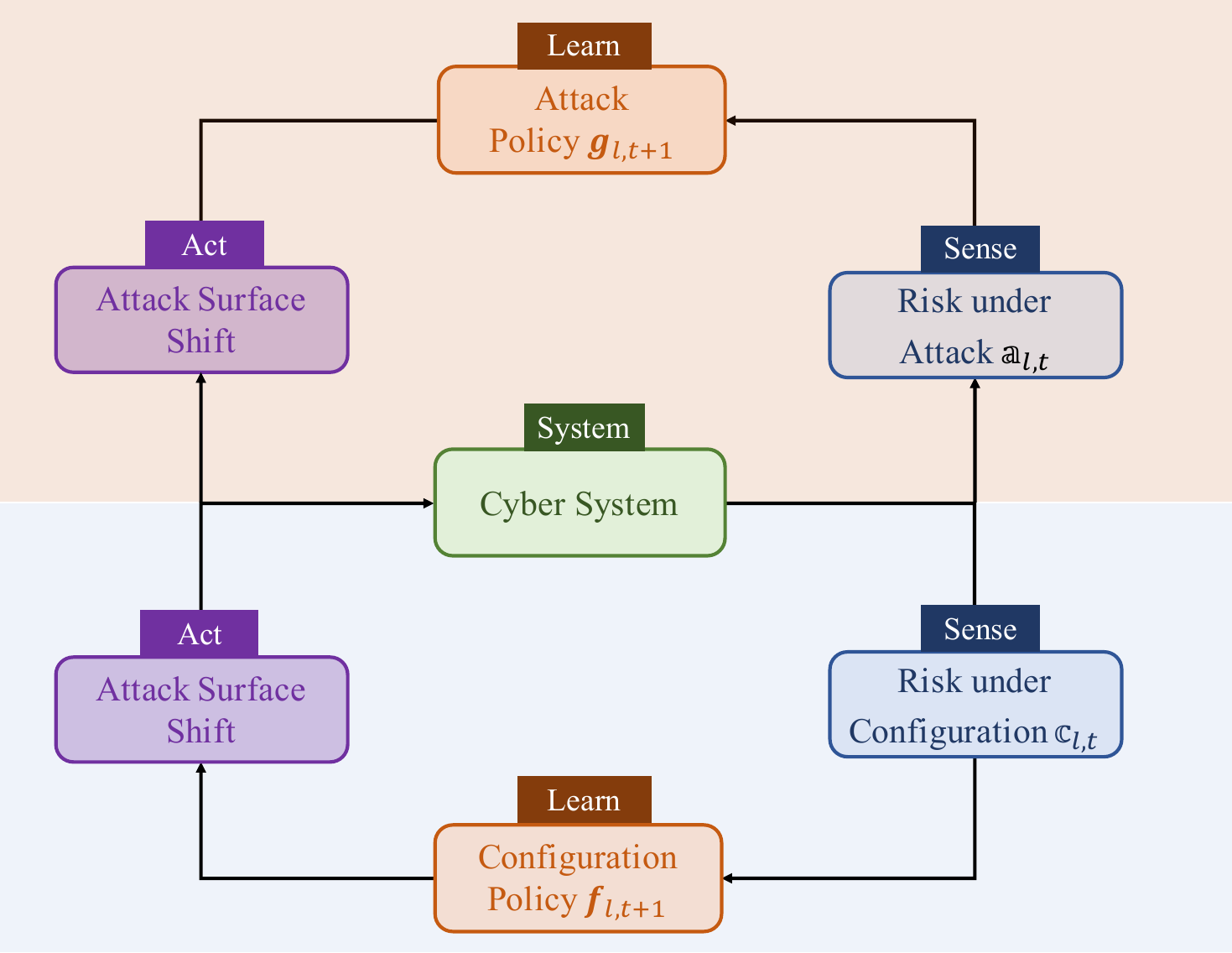}
\caption{ 
Adaptive learning of the multistage MTD game at layer $l$ where the adversarial and defensive learning is presented in red and green, respectively.  
In both players' learning feedback, the procedures of `sense', `learn', and `act' correspond to the boxes of `information acquisition', `decision making', and `security configuration' in Fig. \ref{feedback}, respectively. 
}
\label{fig: Feedback2}
\end{figure}

In practice, the costs $D_{hk}, \forall h\in m_l,k\in n_l$, are unknown and subject to noise. 
Thus, the attacker and the defender need to learn the cost $r_l$ independently during their interaction over time, which is referred to as the procedure of `sense' in Fig. \ref{fig: Feedback2}.  
Then, each player updates his policy based on the estimated cost and then takes an action based on the updated policy, which are referred to as the procedures of  `learn' and `act', respectively, in Fig. \ref{fig: Feedback2}. 

The subscript $t$ denotes the strategy or cost at time $t$. 
Due to the non-cooperative environment, the attacker does not know the configuration and the defender does not know the attack action throughout their interaction. 
Thus, at time $t$, they independently choose actions $\mathbbm{c}_{l,t}\in \mathcal{C}_l$ and $\mathbbm{a}_{l,t}\in \mathcal{A}_l$ according to strategies $\mathbf{f}_{l,t}$ and $\mathbf{g}_{l,t}$, respectively. 
Then, they commonly observe the cost $r_{l,t}$ as an outcome of their action pair $(\mathbbm{c}_{l,t},\mathbbm{a}_{l,t})$. 
Based on the observed cost at time $t$, the defender and the attacker can estimate the average risk $\hat{r}_{l,t}^S: \mathcal{C}_l \rightarrow \mathbb{R}_{+}$ and $\hat{r}_{l,t}^A: \mathcal{A}_l \rightarrow \mathbb{R}_{+}$, respectively, as follows: 
\begin{equation}
\label{eq:utilityLearning}
\begin{split}
& \hat{r}_{l,t+1}^S(c_{l,h})=  \hat{r}_{l,t}^S(c_{l,h}) +\mu_t^S \mathbf{1}_{\{\mathbbm{c}_{l,t}=c_{l,h}\}} (r_{l,t}-\hat{r}_{l,t}^S(c_{l,h})), \forall h\in \{1, \cdots,m_l\},  \\
& \hat{r}_{l,t+1}^A(a_{l,k})=\hat{r}_{l,t}^A(a_{l,k}) +\mu_t^A \mathbf{1}_{\{\mathbbm{a}_{l,t}=a_{l,k}\}} (r_{l,t}-\hat{r}_{l,t}^A(a_{l,k})), \forall k\in \{1, \cdots,n_l\}, 
\end{split}
\end{equation}
where  $\mu_t^S$ and $\mu_t^A$ are the payoff learning rates for the defender and the attacker, respectively. 
The indicators in \eqref{eq:utilityLearning} mean that the defender and the attacker only update the estimate average risk of the observed action $\mathbbm{c}_{l,t}$ and $\mathbbm{a}_{l,t}$ at the current time $t$, respectively. 

The defender uses the estimated risk $\hat{r}_{l,t+1}^S(c_{l,h})$ to update his configuration strategy from $\mathbf{f}_{l,t}$ to $\mathbf{f}_{l,t+1}$. The strategy change involves a reconfigure cost to maneuvering the defense resources and altering the attack surface from $\pi_l(\mathbbm{c}_{l,t})$ to $\pi_l(\mathbbm{c}_{l,t+1})$, where  $\mathbbm{c}_{l,t}$ and $\mathbbm{c}_{l,t+1}$ are selected according to $\mathbf{f}_{l,t}$ and $\mathbf{f}_{l,t+1}$, respectively. 
Define the reconfigure cost as the relative entropy between two consecutive strategies: 
\begin{equation}
R_{l,t}^S:=\sum_{h=1}^{m_l}f_{l,h,t+1}\ln\left(\frac{f_{l,h,t+1}}{f_{l,h,t}}\right). 
\end{equation} 
Then, the reconfigure cost $R_{l,t}^S$ is added to the original expected cost $\sum_{h=1}^{m_l} f_{l,h,t+1} \hat{r}^S_{l,t} (c_{l,h})$ with a parameter $\epsilon_{l,t}^S>0$ to quantify the trade-off between the security and the usability. 
Thus, the defender aims to solve the following optimization problem (SP) at time $t$: 
\begin{equation}
\label{eq:SP}
(\texttt{SP}): \sup_{\mathbf{f}_{l,t+1}\in \Delta \mathcal{C}_l} -\sum_{h=1}^{m_l} f_{l,h,t+1} \hat{r}^S_{l,t} (c_{l,h})-\epsilon_{l,t}^S R_{l,t}^S, 
\end{equation}
which has the following closed-form solution in \eqref{eq:ft-closedform} and the optimal value $W_{l,t}^S$ in \eqref{eq:ft-optimalvalue}. 
\begin{equation}
\label{eq:ft-closedform}
f_{l,h,t+1}=\frac{f_{l,h,t}  e^{ -\frac{\hat{r}_{l,t} (c_{l,h})}{\epsilon_{l,t}^S} } }
{\sum_{h'=1}^{m_l} f_{l,h',t} e^{ -\frac{\hat{r}_{l,t} (c_{l,h'})}{\epsilon_{l,t}^S} }} , \forall h\in \{1,\cdots,m_l\}. 
\end{equation}

\begin{equation}
\label{eq:ft-optimalvalue}
W_{l,t}^S=\epsilon_{l,t}^S \ln \left(\sum_{h=1}^{m_l}  f_{l,h,t}  e^{ -\frac{\hat{r}_{l,t} (c_{l,h})}{\epsilon_{l,t}^S} }  \right). 
\end{equation}

When the value of $\epsilon_{l,t}^S$ is high, the configuration policy changes less and is more usable. However, it is also easier for the attacker to learn the policy and thus reduce the security. 
If $\epsilon_{l,t}^S\rightarrow \infty$, then the configuration policy remains the same $f_{l,h,t+1}=f_{l,h,t}$ and the optimal value in \eqref{eq:ft-optimalvalue} equals $-\sum_{h=1}^{m_l} f_{l,h,t} \hat{r}^S_{l,t} (c_{l,h})$. 
If $\epsilon_{l,t}^S\rightarrow 0$, then the optimal value in \eqref{eq:ft-optimalvalue} equals $\min_{c_{l,h}\in \mathcal{C}_l} \hat{r}_{l,t} (c_{l,h})$. 

Analogously, it takes an attacker time and energy to change the attack strategy to explore new vulnerabilities and exploit them, which leads to the following optimization problem (AP)) for the attacker. 
\begin{equation}
\label{eq:AP}
(\texttt{AP}):  \sup_{\mathbf{g}_{l,t+1}\in \Delta \mathcal{A}_l} -\sum_{k=1}^{n_l} g_{l,k,t+1} \hat{r}^A_{l,t} (a_{l,k})-\epsilon_{l,t}^A \sum_{k=1}^{n_l}g_{l,k,t+1}\ln\left(\frac{g_{l,k,t+1}}{g_{l,k,t}}\right), 
\end{equation}
which has the following closed-form solution in \eqref{eq:gt-closedform} and the optimal value $W_{l,t}^S$ in \eqref{eq:gt-optimalvalue}. 
\begin{equation}
\label{eq:gt-closedform}
    g_{l,k,t+1}=\frac{g_{l,k,t}  e^{ -\frac{\hat{r}_{l,t} (a_{l,k})}{\epsilon_{l,t}^A} } }
{\sum_{k'=1}^{n_l} g_{l,k',t} e^{ -\frac{\hat{r}_{l,t} (a_{l,k'})}{\epsilon_{l,t}^A} }}.  
\end{equation}

\begin{equation}
\label{eq:gt-optimalvalue}
W_{l,t}^A=\epsilon_{l,t}^A \ln \left(\sum_{k=1}^{n_l}  g_{l,k,t}  e^{ -\frac{\hat{r}_{l,t} (a_{l,k})}{\epsilon_{l,t}^A} } \right).  
\end{equation}
The parameter $\epsilon_{l,t}^A>0$ achieves the trade-off between the attacker's benefit of attacks and the cost to learn the most effective attack. 
With all these notations introduced, we use Fig. \ref{fig: Feedback2} to illustrate the attacker's (resp. the defender's) risk learning, policy update, and action implementation in the light orange (resp. light blue) background. 
In either the attacker's adversarial learning or the defender's defensive learning, the learning rule does not depend on the other player's action, yet the observed payoff depends on both players' actions. 

The dynamics for the mixed strategy update (represented by the procedure of `learn' in Fig. \ref{feedback}) in \eqref{eq:ft-closedform} and \eqref{eq:gt-closedform} can be generalized into  the following two learning dynamics \eqref{eq:generalized-ft} and \eqref{eq:generalized-gt}, respectively, with learning rates $\lambda_{l,t}^S,\lambda_{l,t}^A\in [0,1]$.  
\begin{equation}
\label{eq:generalized-ft}
    f_{l,h,t+1}=(1-\lambda_{l,t}^S)f_{l,h,t}+\lambda_{l,t}^S\frac{f_{l,h,t}  e^{ -\frac{\hat{r}_{l,t} (c_{l,h})}{\epsilon_{l,t}^S} } }
{\sum_{h'=1}^{m_l} f_{l,h',t} e^{ -\frac{\hat{r}_{l,t} (c_{l,h'})}{\epsilon_{l,t}^S} }}, \forall h\in \{1,\cdots,m_l\}. 
\end{equation}
\begin{equation}
\label{eq:generalized-gt}
    g_{l,k,t+1}=(1-\lambda_{l,t}^A)g_{l,k,t}+\lambda_{l,t}^A \frac{g_{l,k,t}  e^{ -\frac{\hat{r}_{l,t} (a_{l,k})}{\epsilon_{l,t}^A} } }
{\sum_{k'=1}^{n_l} g_{l,k',t} e^{ -\frac{\hat{r}_{l,t} (a_{l,k'})}{\epsilon_{l,t}^A} }}, \forall k\in \{1,\cdots,n_l\}. 
\end{equation}
If the learning rates $\lambda_{l,t}^S=1$ and $\lambda_{l,t}^A=1$, the learning dynamics \eqref{eq:generalized-ft} and \eqref{eq:generalized-gt} are the same as \eqref{eq:ft-closedform} and \eqref{eq:gt-closedform}, respectively. 
According to the stochastic approximation theory, the learning dynamics \eqref{eq:generalized-ft} and \eqref{eq:generalized-gt} converge to the SPE of the game in \eqref{eq:SPE} under mild conditions \cite{zhu2013game}.

\subsection{Adaptive Honeypot Configuration for Attacker Engagement}
\label{sec:honeypot}

Traditional cybersecurity techniques such as the firewall and intrusion detection systems rely on low-level Indicators of Compromise (IoCs), e.g., hash values, IP addresses, and domain names, to detect attacks. 
However, advanced attacks can revise these low-level IoCs and evade detection. 
Thus, there is an urgent need to disclose high-level IoCs, also referred to as threat intelligence, such as attack tools and Tactics, Techniques, and Procedures (TTP) of the attacker. 
As a promising active cyber defense mechanism, honeypots can gather essential threat intelligence by luring attackers to conduct adversarial behaviors in a controlled and monitored environment. 

\begin{table}[H]
\centering
\caption{Summary of notations for Section \ref{sec:honeypot}. 
\label{table:notation-sec:honeypot}}
{
\begin{tabularx}{\columnwidth}{X l} 
     \hline
\textbf{Variables} &  \textbf{Meaning} \\ \hline
$t\in [0,\infty)$,$k\in \mathbb{Z}_{\geq 0}$ & Index for time and stage \\
$s^k \in \mathcal{S}$ & State at stage $k$\\
$a^k \in \mathcal{A}(s^k)$ & Engagement action\\
$tr$,$z$ & Transition kernel and sojourn distribution \\
$r_1,r_2,r$ & Transition reward, sojourn reward, and investigation reward  \\
$r^{\gamma}$ & Equivalent investigation reward under discounted factor $\gamma\in [0,\infty)$ \\
$\pi\in \Pi:=\mathcal{S}\rightarrow \mathcal{A}$ & Engagement strategy\\
$u(s^0,\pi),v(s^0)$ & Long-term expected utility and value function starting from $s^0\in \mathcal{S}$ \\
\hline
\end{tabularx}
}
\end{table}

\begin{figure}[ht]
\centering
\includegraphics[width=.7 \textwidth]{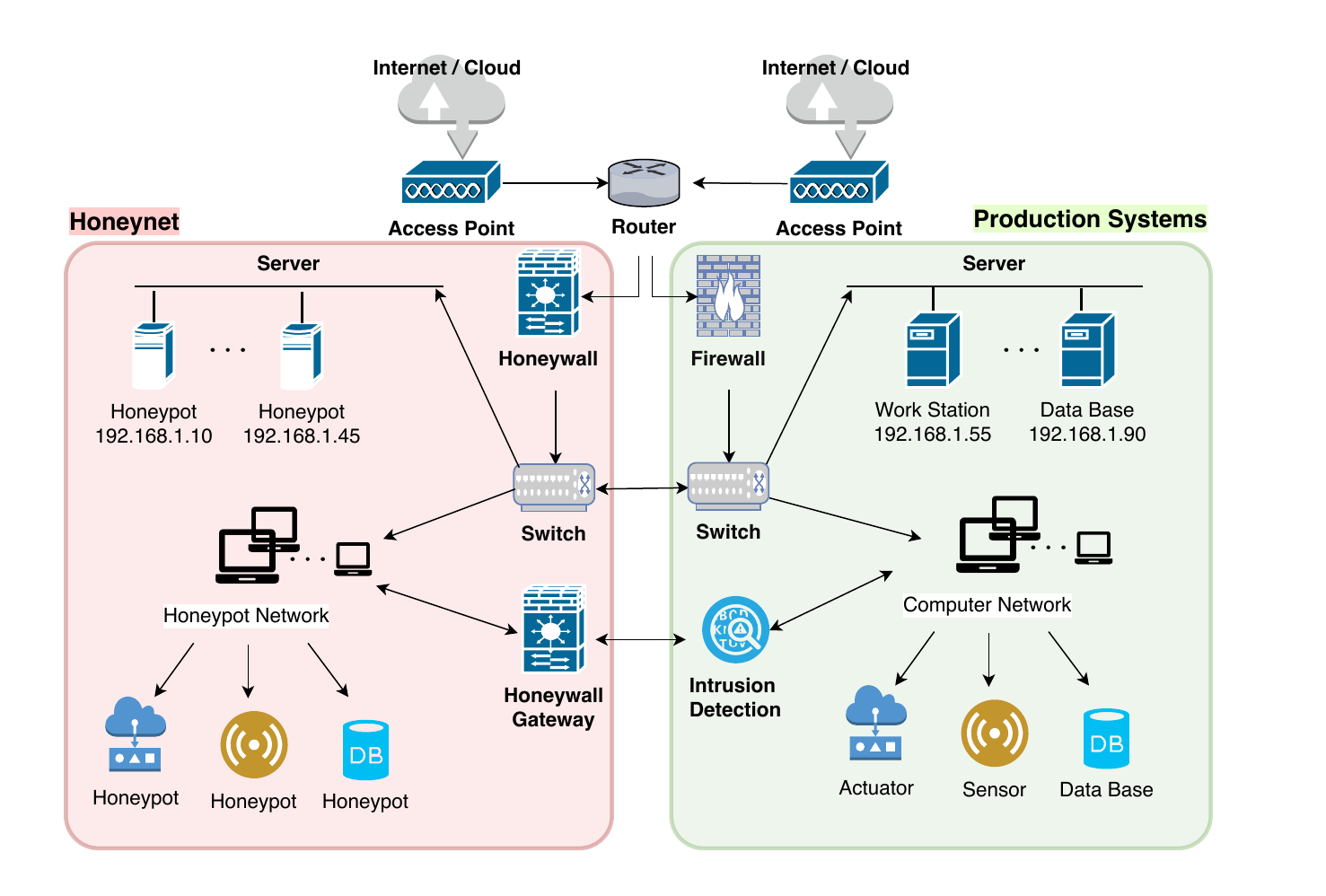}
\caption{
The honeynet in red emulates and shares the same structure as the targeted production system in green. 
 }
 \label{fig: SystemStructure}
\end{figure}

A honeynet is a network of honeypots, which emulates the real production system but has no production activities nor authorized services. 
Thus, an interaction with a honeynet, e.g., unauthorized inbound connections to any honeypot, directly reveals malicious activities.
From an attacker's viewpoint, the production network and the honeypot network share the same structure as shown in Fig. \ref{fig: SystemStructure}. 

An attacker from the production system or external network can be attracted to the honeynet. 
Then, the attacker can move inside the honeynet through either physical (if the connecting nodes are real facilities such as computers) or logical (if the nodes represent integrated systems) links. 
The attacker's transition is restricted by the network topology. 
Once the attack launches an attack at a honeypot node, the defender gains an \textit{investigation reward} by analyzing the attack behaviors and extracting high-level IoCs.  
Since the defender can obtain more threat intelligence when the attack sustains for a longer time, the investigation reward increases with the engaging time. 
The defender aims to dynamically configure the honeynet through proper engagement actions to obtain more threat intelligence while minimize the following three types of risks. 
\begin{itemize}
\item[T1:] Attackers identify the honeynet and thus either terminate on their own or behave misleadingly in honeypots. 
\item[T2:] Attackers circumvent the honeywall and use the honeypots as a pivot to penetrate other production systems. 
\item[T3:] The cost to engage attackers in the honeypots outweighs the resulted investigation reward.  
\end{itemize}

\begin{figure}[ht]
\centering
\includegraphics[width=0.5\textwidth]{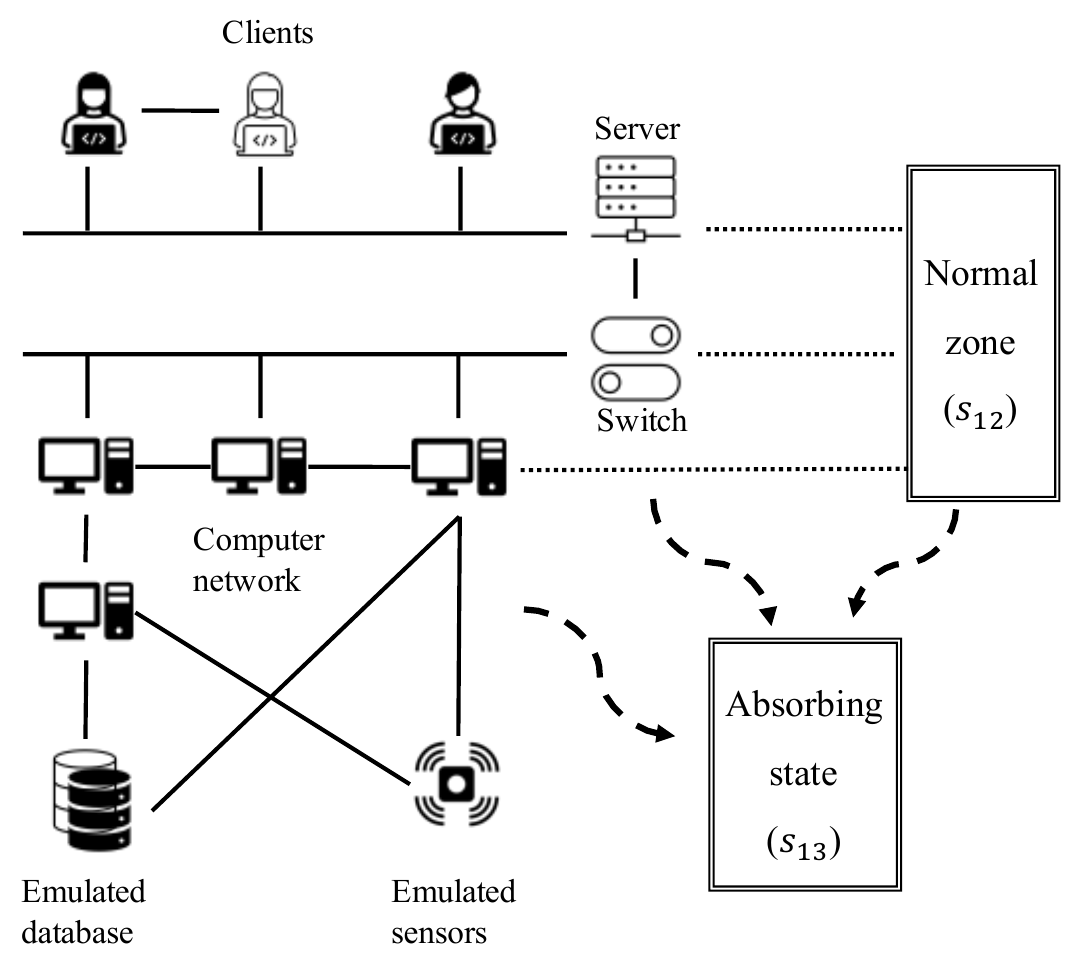}
\caption{
Honeypots emulate different components of the production system.  
Actions $a_E,a_P,a_L,a_H$ are denoted in red, blue, purple, and green, respectively. The size of node $n_i$ represents the state value $v(s_i), i\in \{1,2,\cdots,11\}$.  
}
 \label{fig: SMDPstructure}
\end{figure}

To strike a balance between maximizing the investigation reward resulted and minimizing the risks, {the authors in \cite{huang2019adaptive}} model the attacker's stochastic transition in the honeynet as an infinite-horizon Semi-Markov Decision Process (SMDP) to quantify the long-term reward and risks. 
The SMDP consists of the tuple $ \{t\in [0,\infty), \mathcal{S}, \mathcal{A}({s_j}), tr(s_l| s_j, a_j),  \allowbreak z(\cdot| s_j,a_j,s_l), \allowbreak
r^{\gamma}( s_j,a_j,s_l), \gamma\in [0,\infty)\}$. 
We illustrate each element of the tuple through a $13$-state example in Fig. \ref{fig: SMDPstructure}. 

Each node in Fig. \ref{fig: SMDPstructure} represents a state $s_i\in \mathcal{S}, i\in \{1,2,\cdots,13\}$. 
At time $t\in [0,\infty)$, the attacker is either at one of the honeypot nodes denoted by state $s_i\in \mathcal{S}, i\in \{1,2,\cdots,11\}$, at the normal zone $s_{12}$, or at a virtual absorbing state $s_{13}$ once attackers are ejected or terminate on their own. 
At each state $s_i\in \mathcal{S}$,  the defender can choose an action $a_i\in \mathcal{A}(s_i)$. 
The action set $\mathcal{A}(s_i)$ is finite and depends on the state $s_i\in \mathcal{S}$. 
For example, at  honeypot nodes, the defender can conduct action $a_E$ to eject the attacker, action $a_P$ to purely record the attacker's activities, low-interactive action  $a_L$, or high-interactive action $a_H$, i.e., $\mathcal{A}(s_i):=\{a_E, a_P,a_L,a_H\}, i\in \{1,\cdots, 11\}$. 
At normal zone, the defender can choose either action $a_E$ to eject the attacker immediately, or action $a_A$ to attract the attacker to the honeynet by generating more deceptive inbound and outbound traffics in the honeynet, i.e., $\mathcal{A}(s_{12}):=\{a_E,a_A\}$.  
No actions needed in the virtual absorbing state, i.e., $\mathcal{A}(s_{13}):=\emptyset$. 

Based on the attacker's current state $s_j\in \mathcal{S}$ and the defender's action $a_j\in \mathcal{A}(s_j)$, the attacker transits to state $s_l\in \mathcal{S}$ with probability $tr(s_l|s_j,a_j)$ and the sojourn time at state $s_j$ is a continuous random variable with probability density $z(\cdot| s_j,a_j, s_l)$. 
Once the attacker arrives at a new honeypot $n_i$, the defender dynamically applies an interaction action at honeypot $n_i$ from $\mathcal{A}(s_i)$ and keeps interacting with the attacker until the attacker's next transition. 
Since the defender makes decision at the time of transition, the above continuous-time model over $t\in [0,\infty)$ can be transformed into a discrete decision model at decision epoch $k\in \{0,1,\cdots, \infty\}$. 
The time of the attacker's $k^{th} $ transition is denoted by a random variable $T^k$, the landing state is denoted as $s^k\in \mathcal{S}$, and the adopted action  after arriving at $s^k$ is denoted as $a^k\in \mathcal{A}(s^k)$. 

At decision epoch $k\in \{0,1,\cdots, \infty\}$, the defender's investigation reward $r$ consists an immediate cost $r_1$ of applying engagement action $a^k \in \mathcal{A}(s^k)$ at state $s^k \in \mathcal{S}$ and a reward rate $r_2$, i.e., 
\begin{equation*}
r(s^k,a^k,s^{k+1},T^k,T^{k+1},\tau)=r_1(s^k,a^k,s^{k+1})\mathbf{1}_{\{\tau=0\}}+r_2(s^k,a^k, T^k,T^{k+1},\tau), \tau\in [T^k,T^{k+1}]. 
\end{equation*}
The reward rate $r_2$ at time $\tau\in [T^k,T^{k+1}]$ represents the benefit of threat information acquisition minus the cost rate of persistently generating deceptive traffic. 
Considering a discounted factor of $\gamma\in [0,\infty)$ to penalize the decreasing value of the investigation as time elapses, the defender aims to maximize the long-term expected utility starting from state $s^0$, i.e., 
$
u(s^0,\pi)=\mathbb{E} [\sum_{k=0}^{\infty} \int_{T^k}^{T^{k+1}} e^{-\gamma(\tau+T^k)} (r(S^k,A^k,S^{k+1},T^k,T^{k+1},\tau))d\tau ],  
$
where the \textit{engagement strategy} $\pi\in \Pi$ maps state $s^k\in \mathcal{S}$ to action $a^k\in \mathcal{A}(s^k)$. 
Based on dynamic programming, the value function $v(s^0)=\sup_{\pi \in \Pi }u(s^0,\pi)$ can be represented as 
\begin{align}
\label{eq:DPgeneral}
v(s^0)=\sup_{a^0\in \mathcal{A}(s^0)} \mathbb{E}\left[\int_{T^0}^{T^{1}} e^{-\gamma (\tau+T^0)}r(s^0,a^0,S^{1},T^0,T^{1},\tau)d\tau+e^{-\gamma T^1}v(S^1)\right]. 
\end{align}
Assuming a constant reward rate $r_2(s^k,a^k,T^k,T^{k+1},\tau)=\bar{r}_2(s^k,a^k)$, \eqref{eq:DPgeneral} can be transformed into the equivalent MDP form as follows, i.e., 
\begin{align}
v(s^0)=\sup_{a^0\in \mathcal{A}(s^0)} \sum_{s^1\in \mathcal{S}} tr(s^1|s^0,a^0) (r^{\gamma}(s^0,a^0,s^1)+z^{\gamma}(s^0,a^0,s^1)v(s^1)), \forall s^0\in \mathcal{S}, 
\end{align}
where ${z^{\gamma}}(s^0,a^0,s^1):=\int_0^{\infty}e^{-\gamma \tau} z(\tau|s^0,a^0,s^1)d\tau\in [0,1]$ is the Laplace transform of the sojourn probability density $z(\tau|s^0,a^0,s^1)$ and the equivalent reward 
$r^{\gamma}(s^0,a^0,s^1)\allowbreak
:=r_1(s^0,a^0,s^1)+\frac{\bar{r}_2(s^0,a^0)}{\gamma} (1-z^{\gamma}(s^0,a^0,s^1))\in [-m_c,m_c]$ is assumed to be bounded by a constant $m_c$.

\subsubsection{Optimal Engagement Strategy via Reinforcement Learning}
\begin{figure}
\centering
\includegraphics[width=0.6\textwidth]{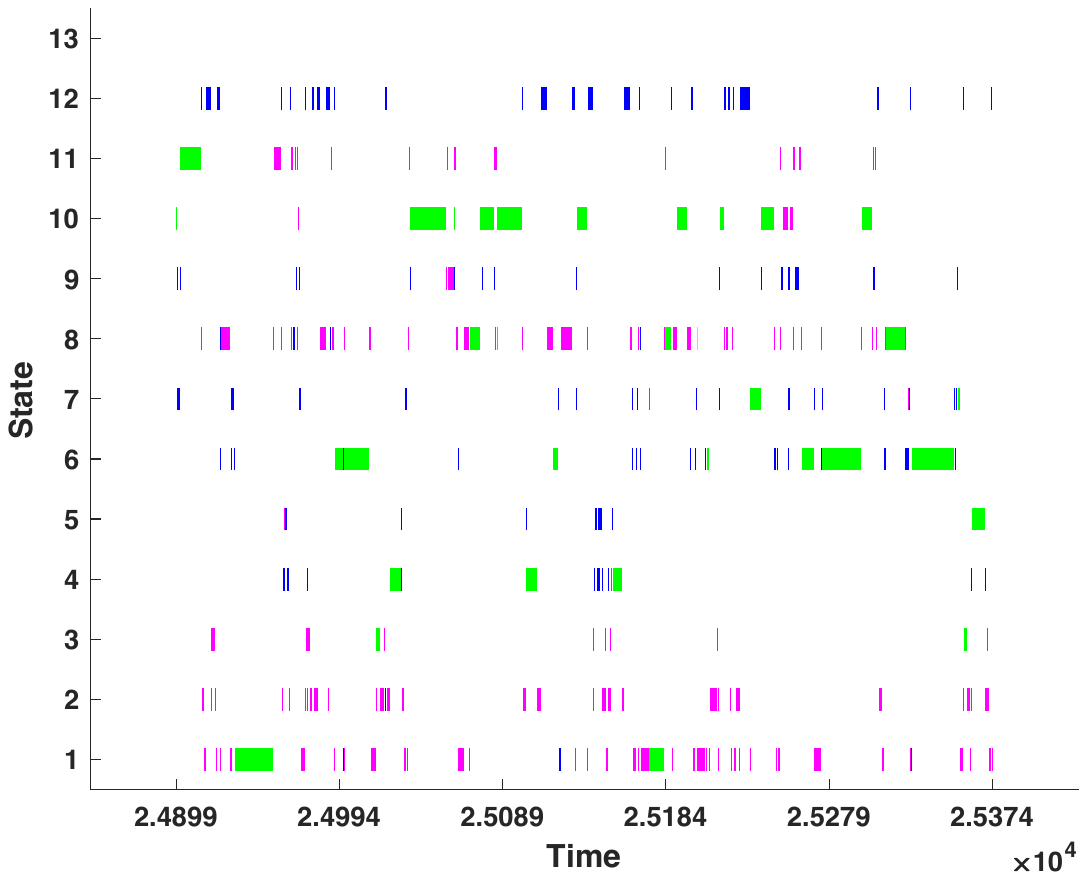}
\caption{
One instance of $Q$-learning of SMDP where the $x$-axis shows the sojourn time and the $y$-axis represents the state transition. 
The chosen actions $a_E,a_P,a_L,a_H$ are denoted in red, blue, purple, and green, respectively. 
 \label{fig: SamplePath}}
\end{figure}

In practice, the defender cannot know the exact SMDP model, i.e., the investigation reward, the attacker's transition probability (and even the network topology), and the sojourn distribution. 
Thus, the defender learns the optimal engagement policy iteratively during the honeynet interaction with the attacker through  the $Q$-learning algorithm for SMDP  \cite{bradtke1995reinforcement}, i.e., 
\begin{equation}
\label{eq:Qlearning}
\begin{split}
Q^{k+1}(s^k,a^k):=&(1-\alpha^k(s^k,a^k))Q^{k}(s^k,a^k)+ \alpha^k(s^k,a^k)[ \bar{r}_1(s^k,a^k,\bar{s}^{k+1})
\\
&
+\bar{r}_2(s^k,a^k)\frac{(1-e^{-\gamma \bar{\tau}^k})}{\gamma}-e^{-\gamma \bar{\tau}^k}\max_{a'\in \mathcal{A}(\bar{s}^{k+1})} Q^k(\bar{s}^{k+1},a')], 
\end{split}
\end{equation}
where $s^k$ is the current state sample,  $a^k$ is the current selected action, $\alpha^k(s^k,a^k)\in (0,1)$ is the learning rate, $\bar{s}^{k+1}$ is the observed state at next stage, $\bar{r}_1,\bar{r}_2$ is the observed investigation rewards, and $\bar{\tau}^k$ is the observed sojourn time at state $s^k$. 
When the learning rate satisfies $\sum_{k=0}^\infty \alpha^k(s^k,a^k)=\infty, \sum_{k=0}^\infty (\alpha^k(s^k,a^k))^2<\infty, \forall s^k\in \mathcal{S}, \forall a^k\in \mathcal{A}(s^k)$, and all state-action pairs are explored infinitely, $\max_{a'\in \mathcal{A}(s^k)} \allowbreak
Q^k(s^k,a'), k\rightarrow \infty$,  in \eqref{eq:Qlearning} converges to value $v(s^k)$ with probability $1$. 
At each decision epoch $k\in \{0,1,\cdots\}$, the action $a^k$ is chosen according to the $\epsilon$-greedy policy, i.e., the defender chooses the optimal action $arg\max_{a'\in \mathcal{A}(s^k)} Q^k(s^k,a')$ with a probability $1-\epsilon$, and a random action with a probability $\epsilon$. 
Fig. \ref{fig: SamplePath} illustrates an exemplary learning trajectory of the state transition and sojourn time under the  $\epsilon$-greedy exploration policy, where the chosen actions $a_E,a_P,a_L,a_H$ are denoted in red, blue, purple, and green, respectively. 

\begin{figure}[ht]
\centering
\includegraphics[width=0.6 \textwidth]{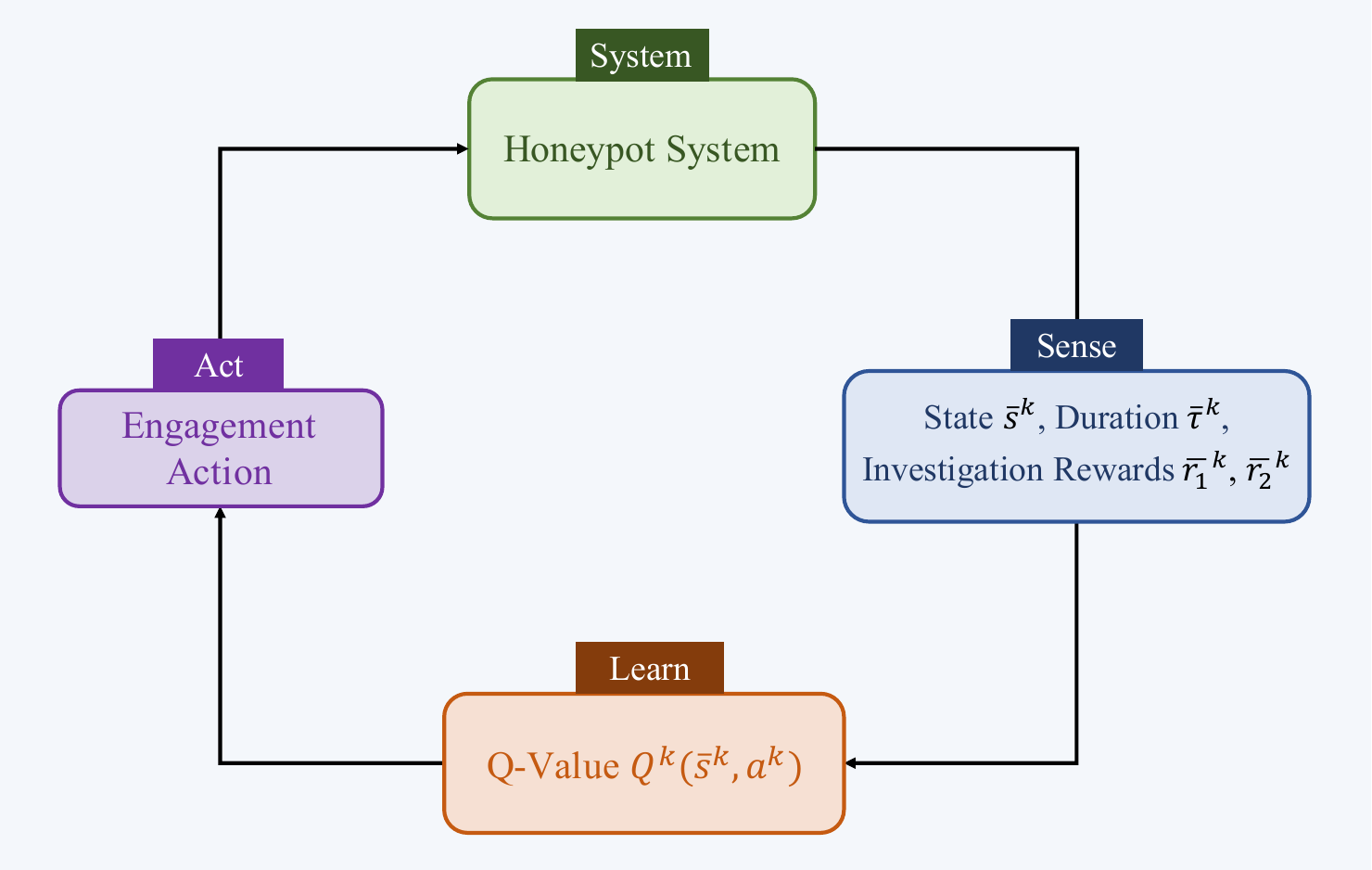}
\caption{
The feedback structure of reinforcement learning methods on SMDP. 
The attacker's characteristics determine the environmental uncertainties and the samples observed in the honeynet. 
}
\label{fig: Feedback3}
\end{figure}

We plot the entire feedback learning structure in Fig. \ref{fig: Feedback3}. 
The environmental uncertainties represented by the red background result from the attacker's behaviors in the honeypots. 
It is assumed that the attacker does not identify the existence of the honeypot. Thus, the observed samples at the procedure of `sense' indicate the attackers' characteristics and contain the correct threat intelligence. 
Then, the defender updates the $Q$-value at each decision epoch $k\in \{0,1,\cdots,\infty\}$ and then chooses the engagement action $a^k$ as shown in the procedures of `learn' and `act', respectively. 

\subsection{Adaptive Alert and Attention Management Strategy against IDoS Attacks}
\label{sec:human}

As an analogy to the DoS attacks that generate a large number of superfluous requests to exhaust the limited computing resources in communication systems, IDoS attacks generate a large number of feints to exhaust the operators' limited cognition resources (e.g., attention, reasoning, and decision-making) in human-in-the-loop systems. 
IDoS attack is a broad class of attacks and poses the following significant security challenges. 
First, the strategical feint generation exerts additional cognition load to the human operators and exacerbates alert fatigue in the age of infobesity. 
Second, IDoS directly target human operators in the Security Operating Center (SOC), which is regarded as the cyber immune system.
Third, it requires operators of a high expertise level to identify feints and respond to alert timely in complicated systems. 
Therefore, there is a need to understand this kind of attentional attack, quantify its consequences and risks, and develop new mitigation methods.

\begin{figure}[ht]
\centering
  \includegraphics[width=.8 \linewidth]{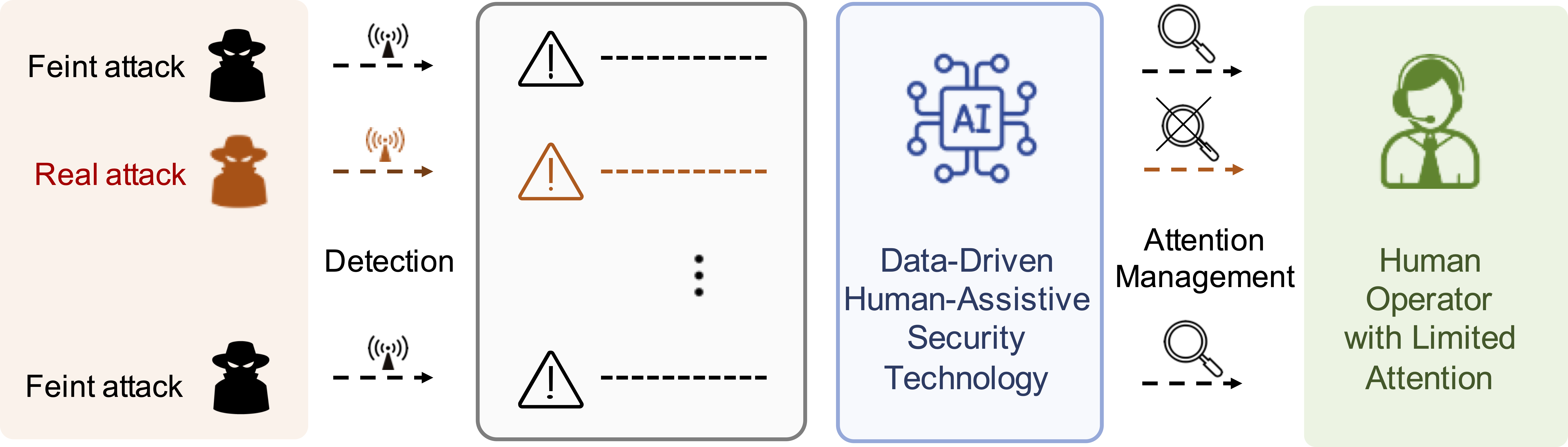}
  \caption{
Interaction among IDoS attacks, human operators, and assistive technologies in orange, green, and blue, respectively. 
  }
\label{fig:AHM}
\end{figure}

\begin{figure}[ht]
\centering
\includegraphics[width=0.85 \textwidth]{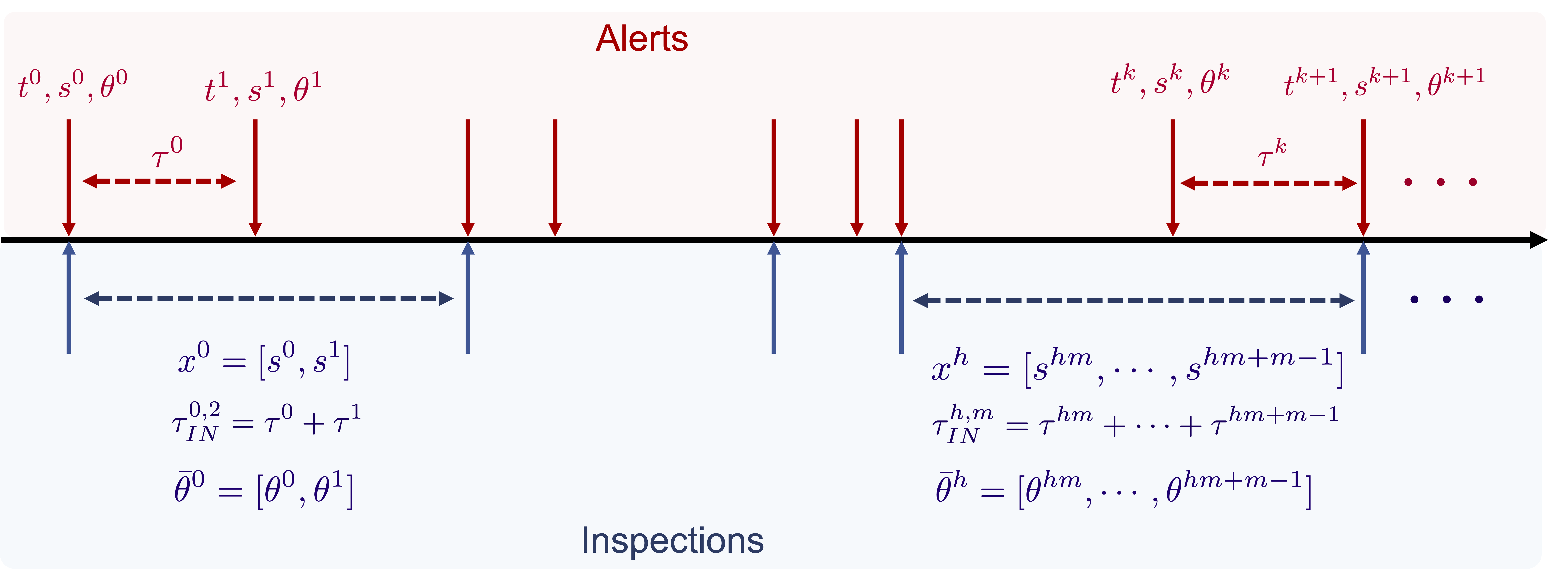}
\caption{ 
The sequential arrival of alerts at \textit{attack stage} $k\in \mathbb{Z}_{\geq 0}$
and 
the periodic manual inspections at \textit{inspection stage} $h\in  \mathbb{Z}_{\geq 0}$ under AM strategy $a_m\in \mathcal{A}$ where $m=2$.  
}
\label{fig:newMDP}
\end{figure}

\begin{table}[H]
\centering
\caption{Summary of notations for Section \ref{sec:human}. 
\label{table:notation-sec:human}}
{
\begin{tabularx}{\columnwidth}{X l} 
     \hline
\textbf{Variables} &  \textbf{Meaning} \\ \hline
$k,h\in \mathbb{Z}_{\geq 0}$ & Index for attack stages and inspection stages \\
$t^k\in [0,\infty)$  &  Arrival time of the $k$-th attack \\
$\tau^k=t^{k+1} - t^k\in [0,\infty)$       &  Time duration between $k$-th and $(k+1)$-th attack\\
$\tau_{IN}^{h,m}:=\sum_{{k}'=hm}^{hm+m-1} \tau^{{k}'}$ & Inspection time at inspection stage $h\in  \mathbb{Z}_{\geq 0}$  \\
$w^k\in \mathcal{W}:=\{w_{FE},w_{RE},w_{UN}\}$       &  Security decision at attack stages $k\in \mathbb{Z}_{\geq 0}$\\
$a_m\in \mathcal{A}$ & Attention management strategy of period $m\in \mathbb{Z}_{>0}$ \\
$\theta^k\in \Theta:=\{\theta_{FE},\theta_{RE}\}$       &  Attack's type at  attack stages $k\in \mathbb{Z}_{\geq 0}$\\
 $\bar{\theta}^{h}:=[\theta^{hm},\cdots,\theta^{hm+m-1}]$  & Consolidated type at inspection stage $h\in  \mathbb{Z}_{\geq 0}$  \\ 
$s^k\in \mathcal{S}$       &  Alert's category label at attack stages $k\in \mathbb{Z}_{\geq 0}$\\
$x^h:=[s^{hm},\cdots,s^{hm+m-1}]$ & Consolidated state at inspection stage $h\in \mathbb{Z}_{\geq 0}$ \\
$c(w^k,s^k;\theta^k),\bar{c}(x^h, a_m; \bar{\theta}^{h})$ & Cost and Consolidated cost \\
$\hat{u}(x^{h},a_m), \bar{u}(s^{hm},a_m)$  & Expected cumulative cost and expected aggregated cumulative cost \\
$ v^{h+1}(\hat{x}^h,a_m),\bar{v}^{h+1}(\hat{s}^{hm},a_m)$ & Learning estimates of ECC and EACC \\
\hline
\end{tabularx}
}
\end{table}
The authors in \cite{RN661} establish a holistic model in Fig. \ref{fig:AHM} to formalize the definition of IDoS attacks, evaluate their severity levels, and assess the induced cyber risks. 
In particular, they model the IDoS attack as stochastic arrivals of feints (denoted by $\theta_{FE}$ with probability $b_{FE}$) and real attacks (denoted by $\theta_{RE}$ with probability $b_{RE}$) characterized by the following Markov renewal process. 
As highlighted by the orange background in Fig. \ref{fig:newMDP}, 
the $k$-th attack, or equivalently the one at \textit{attack stage} $k\in \mathbb{Z}_{\geq 0}$, happen at time $t^k, k\in \mathbb{Z}_{\geq 0}$. 
Let $\tau^k:=t^{k+1}-t^k\in [0,\infty)$ be the inter-arrival time between the $(k+1)$-th attack and the $k$-th attack for all $k\in \mathbb{Z}_{\geq 0}$. 

The alerts triggered by these attacks in real-time cannot directly reveal the \textit{attack's type} denoted by $\theta^k\in \Theta:=\{\theta_{FE},\theta_{RE}\}$ at all attack stages $k\in \mathbb{Z}_{\geq 0}$. 
However, the alerts can provide human operators with a \textit{category label} $s^k\in \mathcal{S}$ based on observable features or traces of the associated attacks at attack stage $k\in \mathbb{Z}_{\geq 0}$. 
Manual inspection of these alerts leads to three \textit{security decisions}: the attack is feint (denoted by $w_{FE}$), the attack is real (denoted by $w_{RE}$), or the attack's type is unknown (denoted by $w_{UN}$). 
Let $w^k\in \mathcal{W}:=\{w_{FE},w_{RE},w_{UN}\}$ denote the human operator's security decision of the $k$-th alert. 
Since each human operator has limited attention, frequent alert pop-ups can distract humans from the current alert inspection and result in alert fatigue. 
To compensate for the human's attention limitation, one strategy is to intentionally make some alerts less noticeable, e.g.,  without sounds or in a light color. 
Then, the human can pay sustained attention to the alert currently under inspection. 
Those inconspicuous alerts receive the security decision $w_{UN}$ and are assigned to other available inspectors with an additional cost of human resources and inspection delays. 

The authors focus on the class of periodic \textit{Attention Management (AM) strategies}, denoted by $\mathcal{A}:=\{a_m\}_{m\in \mathbb{Z}_{>0}}$ and assume that the human operator can only notice and inspect an alert when it is highlighted. 
Then, AM strategy $a_m\in \mathcal{A}$ means that the human operator inspects the alerts at attack stages $k=hm, h\in  \mathbb{Z}_{\geq 0}$. 
The attack stages during the $h$-th inspection are referred to as the \textit{inspection stage} $h\in  \mathbb{Z}_{\geq 0}$. 
Then, under AM strategy $a_m\in \mathcal{A}$, each inspection stage contains $m$ attack stages as shown in the blue background of Fig. \ref{fig:newMDP} and the $h$-th inspection has a duration of $\tau_{IN}^{h,m}:=\sum_{{k}'=hm}^{hm+m-1} \tau^{{k}'}$ for all $h\in  \mathbb{Z}_{\geq 0}$. 
The \textit{consolidated state} $x^h:=[s^{hm},\cdots,s^{hm+m-1}]\in \mathcal{X}:= \mathcal{S}^m$ and the \textit{consolidated type} $\bar{\theta}^{h}:=[\theta^{hm},\cdots,\theta^{hm+m-1}]\in \bar{\Theta}:= \Theta^m $ consist of the category labels and types of $m$ successive alerts, respectively, at inspection stage $h\in  \mathbb{Z}_{\geq 0}$.

Denote $c(w^k,s^k;\theta^k)$ as the operator's cost at attack stage $k\in \mathbb{Z}_{\geq 0}$ when the alert's category label is $s^k\in \mathcal{S}$, the attack's type is $\theta^k\in \Theta$, and the security decision is $w^k\in \mathcal{W}$. 
At attack stages where alerts are inconspicuous, i.e., for all $k\neq hm, h\in  \mathbb{Z}_{\geq 0}$, the security decision is  $w_{UN}$ without manual inspection, which incurs an \textit{uncertainty cost}  $c(w_{UN}, s^k;\theta^k)>0$.  
At attack stages of highlighted alerts, i.e., for all $k=hm, h\in  \mathbb{Z}_{\geq 0}$, 
the human operator obtains a reward (resp. cost) for correct (resp. incorrect) security decisions. 
If the human operator remains uncertain about the attack's type after the inspection time $\tau_{IN}^{h,m}$, i.e., $w^{hm}=w_{UN}$, there is the uncertainty cost $c(w_{UN}, s^{hm};\theta^{hm})$. 
Define the human operator's \textit{consolidated cost} at inspection stage $h\in  \mathbb{Z}_{\geq 0}$ as 
\begin{equation}
\label{eq:consolidated cost}
        \bar{c}(x^h, a_m; \bar{\theta}^{h}) :=(m-1) c(w_{UN}, s^{hm};\theta^{hm}) + \sum_{w^{hm}\in \mathcal{W}}  \Pr(w^{hm}|x^h,a_m;\bar{\theta}^{h})  c(w^{hm},s^{hm};\theta^{hm}),   
\end{equation}
where the \textit{decision probability} $\Pr(w^k|s^k, a_m;\theta^k)$ represents the probability of human making decision $w^k\in \mathcal{W}$ when the attack's type is $\theta^k\in \Theta$, the category label is $s^k\in \mathcal{S}$, and the AM strategy is $a_m\in \mathcal{A}$. 
With discounted factor $\gamma\in (0,1)$, define $
\hat{u}(x^{h},a_m) := \mathbb{E} [ \sum_{h=h_0}^{\infty}  (\gamma)^h \cdot \bar{c}(x^{h},a_m;\bar{\theta}^{h}) ]$ as the \textit{Expected Cumulative Cost (ECC)} under $x^{h}$ and $a_m$. 
To reduce the dimension of the samples and enable evaluations with no delay, the authors further define the \textit{Expected Aggregated Cumulative Cost (EACC)} as 
\begin{equation}
\label{def:ACC}
\begin{split}
        \bar{u}(s^{hm},a_m):=\sum_{s^{hm+1}, \cdots, s^{hm+m-1}\in\mathcal{S}} \bigg[ \Pr(s^{hm+1}, \cdots, s^{hm+m-1}|s^{hm}) \cdot \hat{u}([s^{hm}, \cdots, s^{hm+m-1}],a_m) \bigg], 
\end{split}
\end{equation}
and prove the \textit{computational equivalency} between ECC and EACC under mild conditions. 
Both ECC $\hat{u}(x^0,a_m)$ and EACC $\bar{u}^0(s^0,a_m)$ evaluate the long-term performance of the AM strategy $a_m\in \mathcal{A}$ on average. However, EACC depends on $s^{hm}$ but not on $s^{hm+1}, \cdots, s^{hm+m-1}$. 
Thus, ECC and EACC are referred to as the \textit{consolidated} and the \textit{aggregated} IDoS risks, respectively.

\subsubsection{Optimal Attention Management Strategy via Reinforcement Learning}
Since the parameters of the Markov renewal process are usually unknown, Temporal-Difference (TD) learning are used to evaluate the performance of the AM strategy $a_m\in \mathcal{A}$ based on the inspection results in real-time. 
Define  $v^h(x^h,a_m)$ as the estimated value of $\hat{u}(x^h,a_m)$ at the inspection stage $h\in  \mathbb{Z}_{\geq 0}$, then
\begin{equation}
\label{eq:TDgeneral}
    v^{h+1}(\hat{x}^h,a_m)=(1-\alpha^h(\hat{x}^h) ) v^{h}(\hat{x}^h,a_m)+ \alpha^h(\hat{x}^h) (\hat{c}^h+\gamma v^{h}(\hat{x}^{h+1},a_m) ), 
\end{equation}
where $\hat{x}^h$ (resp. $\hat{x}^{h+1}$) is the observed state value at the current inspection stage $h$ (resp. the next inspection stage $h+1$), $\alpha^h(\hat{x}^h)\in (0,1)$ is the learning rate, and $\hat{c}^h$ is the observed cost at stage $h\in  \mathbb{Z}_{\geq 0}$. 
Alternatively, define $\bar{v}^h(x^h,a_m)$ be the estimated value of $\bar{u}(s^{hm},a_m)$ at the inspection stage $h\in  \mathbb{Z}_{\geq 0}$, then
\begin{equation}
\label{eq:TDspecial}
    \bar{v}^{h+1}(\hat{s}^{hm},a_m)=(1-\bar{\alpha}^h(\hat{s}^{hm}) ) v^{h}(\hat{s}^{hm},a_m)+ \bar{\alpha}^h(\hat{s}^{hm}) (\hat{c}^h+\gamma \bar{v}^{h}(\hat{s}^{(h+1)m},a_m) ), 
\end{equation}
where $\hat{s}^{hm}$ (resp. $\hat{s}^{(h+1)m}$) is the observed state value at the current inspection stage $h$ (resp. the next inspection stage $h+1$), $\bar{\alpha}^h(\hat{s}^{hm})\in (0,1)$ is the learning rate, and $\hat{c}^h$ is the observed cost at stage $h\in  \mathbb{Z}_{\geq 0}$. 

\begin{figure}[ht]
    \centering 
\begin{subfigure}{0.45\textwidth}
  \includegraphics[width=\linewidth]{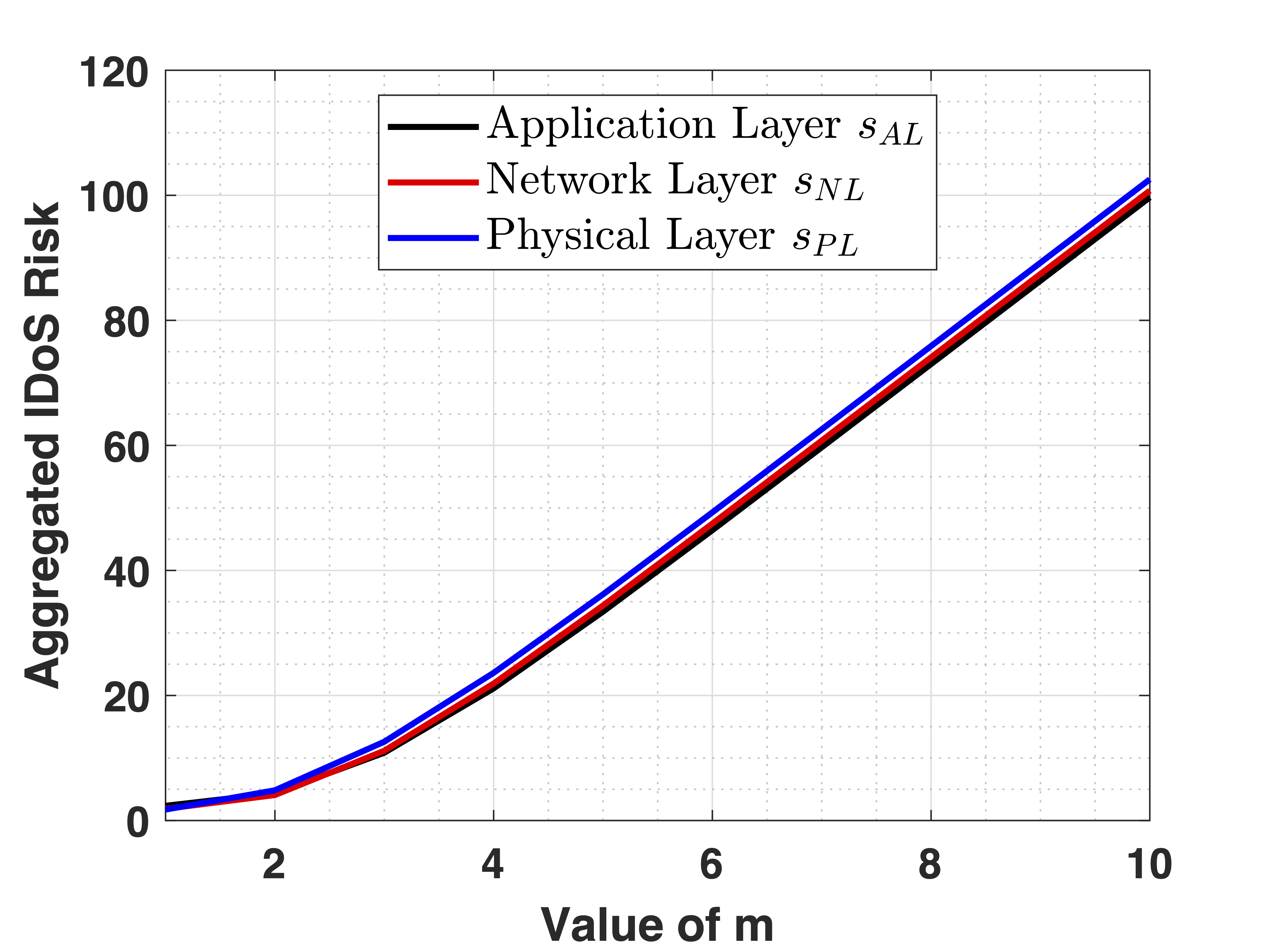}
  \caption{\label{fig:AggregatedriskVSmcUN20} 
High cost $c(w_{UN}, s^{hm};\theta^{hm})=20,\forall s^{hm}, \theta^{hm}$. 
   }
\end{subfigure}\hfil 
\begin{subfigure}{0.45\textwidth} 
  \includegraphics[width=\linewidth]{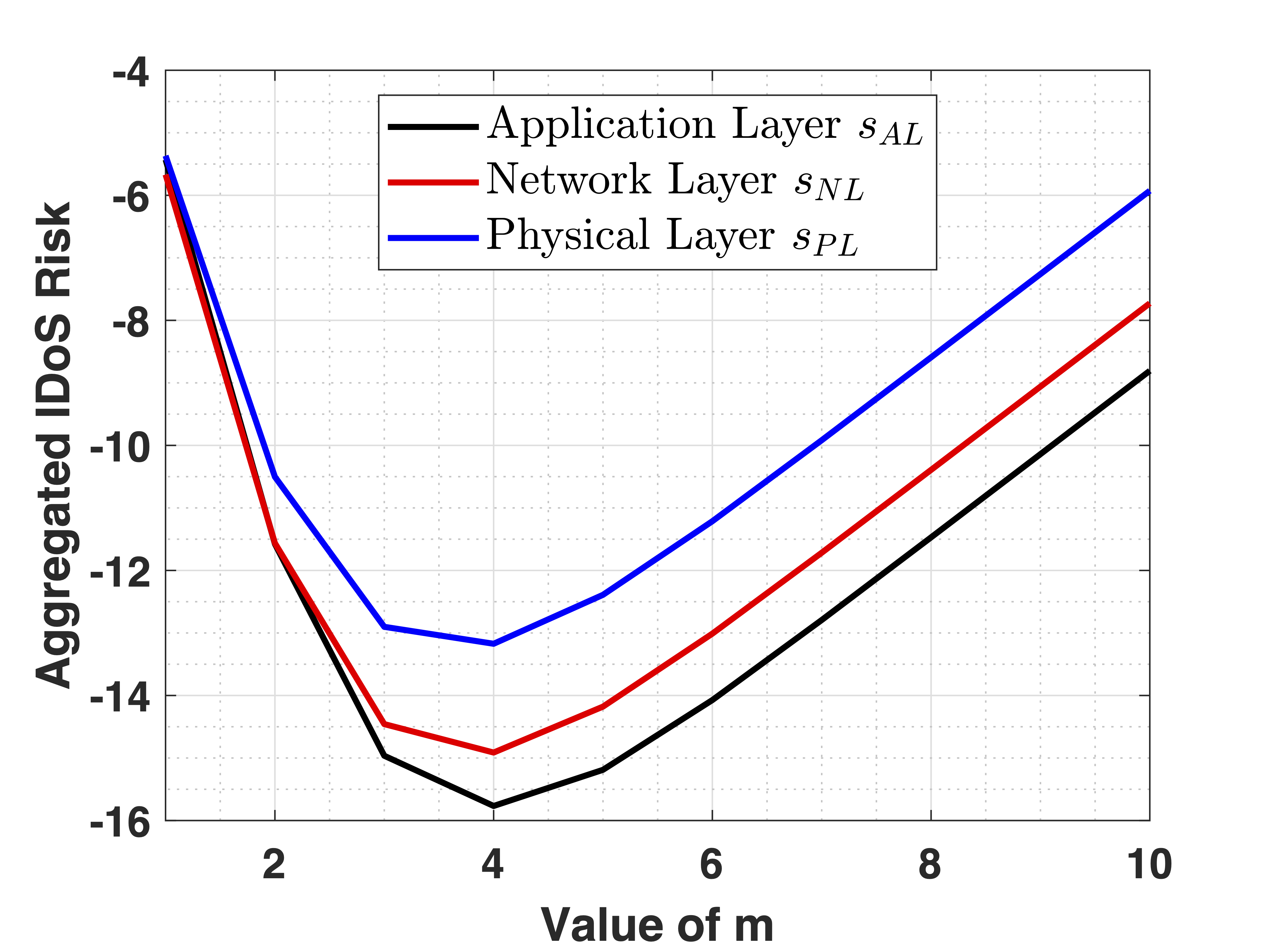}
  \caption{\label{fig:AggregatedriskVSmcUN02}
Low cost $c(w_{UN}, s^{hm};\theta^{hm})=0.2,\forall s^{hm}, \theta^{hm}$. 
 }
\end{subfigure}\hfil 
\caption{
Aggregated IDoS risks under $s_{AL}$, $s_{NL}$, and $s_{PL}$ in black, red, and blue. A small $m$ represents a coarse inspection with a large number of alerts while a large $m$ represents a fine inspection of a small number of alerts. 
}
\label{fig:withAM2}
\end{figure}

Finally, the authors provide a numerical case study where category labels $s_{AL}$, $s_{NL}$, and $s_{PL}$ in the set $\mathcal{S}=\{s_{AL},s_{NL},s_{PL}\}$ represent the application layer, network layer, and physical layer, respectively. 
Fig. \ref{fig:AggregatedriskVSmcUN20} and Fig. \ref{fig:AggregatedriskVSmcUN02} illustrate the impact of high and low uncertainty costs $c(w_{UN}, s^{hm};\theta^{hm})$, respectively. 
If the uncertainty cost is much higher than the expected reward of correct decision-making, then the detailed inspection and correct security decisions are not of priority. 
As a result, the `spray and pray' strategy should be adopted; i.e., let the operator inspect as many alerts as possible and use the high quantity to compensate for the low quality of these inspections.  
Under this scenario, $\bar{u}(s^{hm},a_m)$ increases with $m\in \mathbb{Z}_{>0}$ for all $s^{hm}\in \mathcal{S}$ as shown in Fig. \ref{fig:AggregatedriskVSmcUN20}. 
If the uncertainty cost is of the same order as the inspection reward on average, then increasing $m$ in a certain range (e.g., $m\in\{1,2,3,4,\}$ in Fig. \ref{fig:AggregatedriskVSmcUN02}) can increase the probability of correct decision-making and reduce the aggregated IDoS risk. 
The loss of alert omissions outweighs the gain of detailed inspection when $m$ is beyond that range. 

\begin{figure}[h]
\centering
\includegraphics[width=.6 \columnwidth]{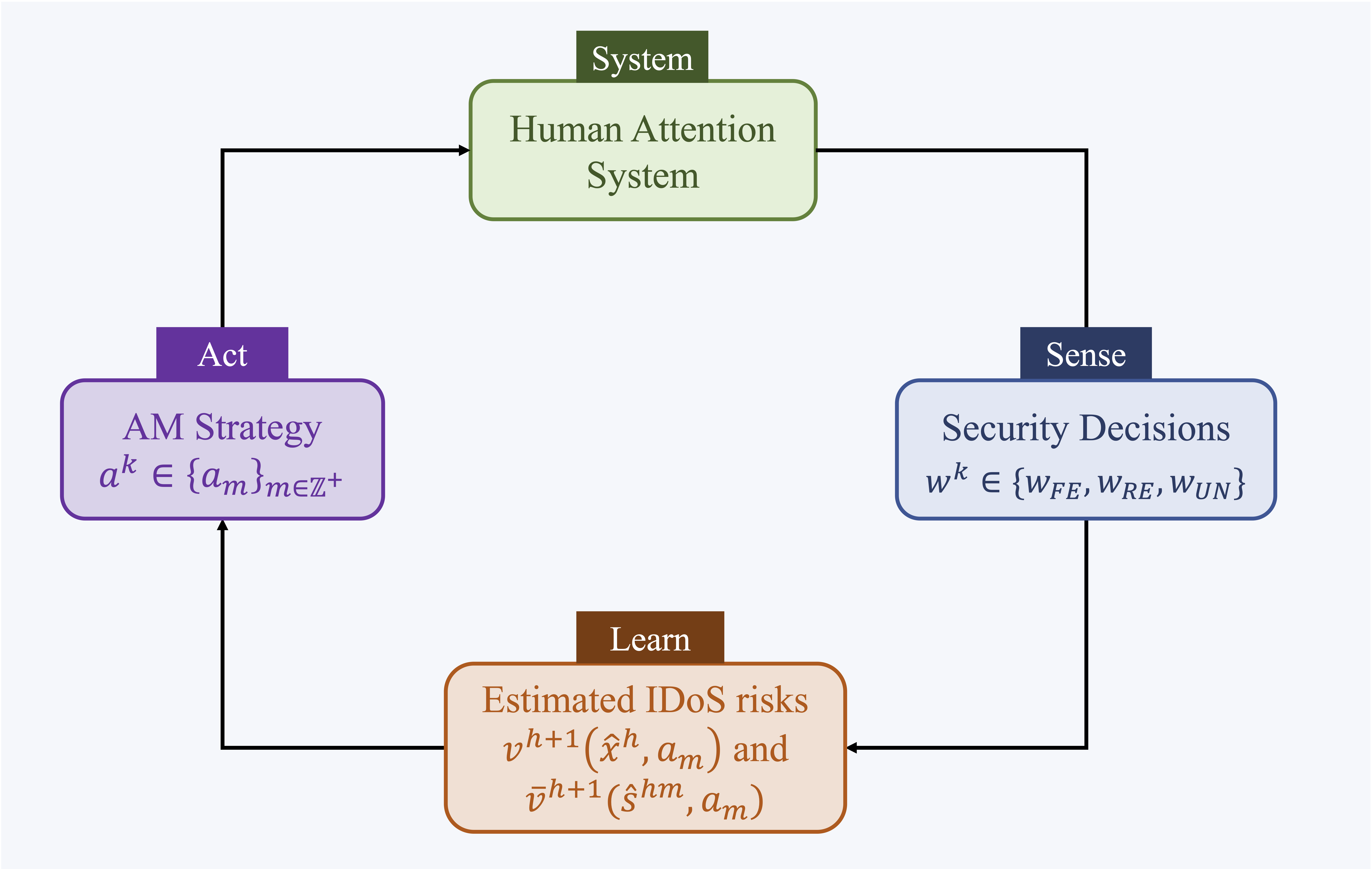} 
\caption{ 
The adaptive learning loop of the optimal alert and attention management strategy to combat IDoS attacks. 
}
\label{fig:feedbackhuman}
\end{figure}

As an analogy to Fig. \ref{fig: Feedback2} and \ref{fig: Feedback3}, we visualize the RL feedback in Fig. \ref{fig:feedbackhuman}. 
Different from the previous two, the cyber system is replaced by the human operator's attention system for alert inspections and responses. 

\section{Reinforcement Learning in Adversarial Environment}\label{sec:RLAdvEnv}

As more and more networked systems are equipped with RL techniques to improve their performance, security, and resilience, the application of RL also creates opportunities for malicious third parties to exploit. Hence, it is of equal importance to study the security problems of RL itself than RL-enabled networked systems.  The successful implement of RL relies on accurate and consistent feedback from the environment, which can be easily guaranteed in a simulated environment. However, in practice, especially in the presence of adversarial interventions, accurate and consistent feedback from the environment is unlikely to be guaranteed. For example, adversaries can manipulate reward signals by performing data injection attacks and prevent an agent from receiving reward signals by launching DoS attacks on the communication channel. Without consistent and/or accurate feedback from the environment, the RL algorithm can either fail to learn a good strategy or be tricked into a 'trap' policy that the attacker aims for. The failure of RL algorithms under adversarial interventions can lead to a catastrophe when RL is applied in critical domains. For example, self-driving platooning vehicles can collide with each other when the measurement data is manipulated \cite{behzadan2019adversarial}; drones equipped with RL techniques can be weaponized by terrorists to create chaotic and vicious situations where they are commanded to crash into a crowd or a building \cite{huang2019deceptive}.

Hence, it is imperative to study RL under an adversarial environment. In the last five years, there is a surge in terms of the number of papers that studies the security issues of RL \cite{huang2019deceptive, wang2020reinforcement, huang2020manipulating,zhang2020adaptive,rakhsha2020policy,chen2021adversarial,sun2020vulnerability,sun2021strongest,banihashem2021defense,liu2021deceptive,ma2019policy,xu2021transferable,figura2021adversarial,majadas2021disturbing}. The vulnerabilities of RL comes from the information exchange between the agent and the environment. As demonstrated in Fig. \ref{feedback}, the agent receives reward signals and state observations from the environment and sends action protocols to the environment. Suppose the agent is conducting real-time learning by communicating with the environment. In that case, the RL agent is subject to DoS attacks, jamming attacks, spoofing attacks, and data injection attacks that all networked systems might suffer. If the agent learns from batches of stored data from previous experience, the RL agent might suffer from data poisoning attacks, test-item attacks, etc. To understand RL under an adversarial environment, first, we need to understand the adversarial behaviors of the attackers. We need to develope an attack mode that characterizes the attacker's objective, the available attacks, the limitation of attacks, and the information available to the attacker. The second is to understand how the specific attacks on the RL algorithms impact the learning results. To do so, we need to develop metrics that measure the success of the attacks, and the consequences rendered by the attacks. With the understanding of the attack model and its impact on the learning results, the last is to design defense mechanisms to protect RL algorithms from being degenerated. The defensive acts may include the detection and removal of corrupted feedback, resilient RL in the absence of feedback signals, and deploying cryptography techniques to ensure confidentiality, etc. In this section, we discuss the security problems faced by RL and review relevant works based on the three possible signals that the attacker can target: the reward, the sensing, and the actuating signals.

\subsection{Attacks on the Reward}\label{subsec:attacksReward}

Reward/cost are the most important signals that the RL agent absorbs in order to learn a proper strategy \cite{sutton2018reinforcement}. The most direct way, perhaps also the easiest way, to trick the RL agent into learning a nefarious strategy is to falsify the reward/cost signals or poison reward/cost data. Huang et al. show that the attacker can alter the learned strategy of many states by poisoning the reward in one state \cite{huang2019deceptive}. In the past two years, many research works have been focused on the reward-poisoning attacks on RL \cite{huang2019deceptive, wang2020reinforcement,huang2020manipulating, zhang2020adaptive, banihashem2021defense,ma2019policy,majadas2021disturbing}. In \cite{wang2020reinforcement}, Wang et al. have investigated the effect of perturbed rewards on a list of RL algorithms. The rewards received by the RL agent are perturbed with a certain probability, and the rewards take values only on a finite set. Here, the rewards are unintentionally perturbed, meaning this work focuses on a robust perspective other than a security perspective. 

Huang and Zhu \cite{huang2019deceptive} have studied the security perspective of cost manipulation in RL. In this work, the attack model is described by the information available to the attacker, the attacer's actions, and his/her objectives.
\begin{itemize}
\item \textit{Information the attacker knows}: The authors consider an omniscient attacker who knows almost all information regarding the learning process.
\item \textit{Actions available to the attacker:} The attacker can manipulate the cost signals in certain states. Formally, let $\tilde{c} \in\mathcal{S} \times \mathcal{A} \rightarrow \mathbb{R}_+$ be the cost function manipulated by the attacker. We say the attacker can only alter the cost signals at states $\mathcal{S}^\dagger \subset \mathcal{S}$, then we have $\tilde{c}(s,a) = c(s,a)$ for all $s\in \mathcal{S} \backslash \mathcal{S}^\dag, a\in\mathcal{A}$, meaning the attacker cannot manipulate the cost signals at states other than the ones in $\mathcal{S}^\dag$. Here, $c$ is the original true cost function. The authors also consider other types of constraints to describe the limitation of the attacker's power, e.g., the manipulated cost is bounded, i.e., $\Vert c-\tilde{c} \Vert \leq B$, where $\Vert \cdot \Vert$ is an appropriate norm.

\begin{table}[H]
\centering
\caption{Summary of notations for Section \ref{subsec:attacksReward}. 
\label{table:notation-subsec:attacksReward}}
{
\begin{tabularx}{\columnwidth}{X l} 
     \hline
\textbf{Variables} &  \textbf{Meaning} \\ \hline
$\mathcal{S},\mathcal{A}$ & State space, action Space \\
$\tilde{c}$ & Manipulated cost function\\
$\mathcal{S}^\dag$ & Set of states where the attacker can manipulate the corresponding cost signals\\
$B$ & The bound of the manipulation\\
$\pi^\dag$ & A nefarious strategy learned from manipulated signals \\
$\pi^*$ & A optimal policy learned from accurate signals\\
$\beta$ & Discount factor\\
$\tilde{Q}^*$ & Optimal $Q$ function learned from manipulated signals\\
$r_t^0$ & True reward signal at time $t$\\
$D\coloneqq(s_t,a_t,r_t^0,s_t')$ & Training set\\
$\mathbf{r}\coloneqq (r_0,\cdots,r_T)$ & Manipulated rewards\\
$\hat{R}$ & Constructed reward function using the training set\\
$\hat{P}$ & Constructed transition probabilities using the training set\\
\hline
\end{tabularx}
}
\end{table}

\item \textit{The objective of the attacker:} The objective of the attacker is to trick the RL agent into learning a nefarious strategy $\pi^\dag$ instead of the original optimal strategy $\pi^*$.
\end{itemize}

Among many RL algorithms, Huang and Zhu have studied $Q$-learning algorithm and its limiting behavior under the manipulated cost signals. The authors develop some fundamental limits regarding the attack model. We highlight a few results in this section. 

\begin{theorem}[\cite{huang2019deceptive}]\label{theo:lipschitz}
Let $Q^*$ denote the $Q$-factor learned from $Q$ learning algorithm (\ref{eq:QAlgo}) with the true cost signals $c$ and $\tilde{Q}^*$ be the $Q$-factor learned from the manipulated cost signals $\tilde{c}$. Then, we have
\begin{equation}\label{eq:lipschitz}
\Vert \tilde{Q}^* - Q^* \Vert \leq \frac{1}{1-\beta} \Vert \tilde{c} - c \Vert,
\end{equation}
where $\beta$ is the discount factor of the underlying MDP problem.
\end{theorem}

Theorem \ref{theo:lipschitz} states that manipulation on cost $c$ using a small perturbation does not cause significant changes in the limit point of the Q learning algorithm. This result indicates that the attacker cannot cause a significant change in the learned $Q$-factor by just a tiny perturbation in the cost signals. This is a feature known as stability observed in problems that possess contraction mapping properties. One conclusion we can make from (\ref{eq:lipschitz}) is that if the attacker wants to successfully make the RL agent learn the nefarious strategy $\pi^\dag$, the attacker has to manipulate the cost signals with magnitude more significant than a certain value:
$$
\Vert \tilde{c} - c\Vert  \geq (1-\beta) \inf_{Q\in\mathcal{V}_{\pi^\dag}} \Vert Q- Q^*  \Vert,
$$
where $\mathcal{V}_{\pi}= \{Q\in \mathbb{R}^{|\mathcal{S}| \times |\mathcal{A}|}: Q(s,\pi(s))< Q(s,a), \forall s\in\mathcal{S}, \forall a \neq \pi(s)  \}.$

Huang and Zhu have also shown that to make the agent learn the nefarious strategy $\pi^\dag$, the manipulated cost signal has to satisfies 
\begin{equation}\label{eq:cost_conds}
\tilde{c}(s, a) > (\mathbbm{1}_s - \beta P_{sa})^T(1- \beta P_{\pi^\dag})^{-1} \tilde{c}_{\pi^\dag},
\end{equation}
where $\mathbbm{1}_s \in \mathbb{R}^{|\mathcal{S}|}$ is a $|\mathcal{S}|$-dimensional vector whose elements are all zero except that the $s$-th element is $1$, $P_{sa} = (p(s,1,a), p(s,2,a),\cdots, p(s,|\mathcal{S}|,a))^T \in \mathbb{R}^{|\mathcal{S}|}$, and $P_{\pi} \in \mathbb{R}^{|\mathcal{S}| \times |\mathcal{A}|}$ is a matrix whose $ij$-component is $[P_{\pi}]_{i,j} = p(i,j,\pi(i))$.  Here, $p(s,s',a)$ is the probability that given current state $s$ and current action $a$, the next state is $s'$, which characterizes the transition kernel of the underlying MDP problem. If the attacker manipulates the cost signals such that (\ref{eq:cost_conds}) is satisfied, the agent will be misled into the nefarious strategy $\pi^\dag$. The conditions in (\ref{eq:cost_conds}) can then be utilized by the attacker to design an optimal manipulation of the cost signals that optimizes the cost of manipulating the cost signals incurred to the attack. Huang and Zhu have also shown theoretically and numerically that the attacker can achieve his/her objective by manipulating the cost signals in a small subset of states.

In \cite{ma2019policy}, Ma et al. have studied a similar problem for model-based RL. The attacker poisons the reward signals stored in a dataset that is used to train an RL agent. The attack model is given as
\begin{itemize}
\item \textit{Information the attacker knows}: The attacker has access to the training set $D = (s_t,a_t, r^0_t,s'_t)_{t=0:T-1}$, in which $r^0_t$ is the true reward signal the agent has at time $t$. The attacker also knows the model-based RL learner's algorithms.

\item \textit{Actions available to the attacker:} The attacker is able to arbitrarily manipulate the rewards $\mathbf{r}^0 = (r^0_0,\cdots, r^0_{T-1})$ in $D$ into $\mathbf{{r}} = ({r}_0,\cdots,{r}_T)$.

\item \textit{The objective of the attacker:} The objective of the attacker is to mislead the RL agent to learn a nefarious strategy $\pi^\dag$ while minimizing his/her attack cost $\Vert \mathbf{r}^0 - \mathbf{{r}} \Vert$.
\end{itemize}

With the attack model being defined as above, Ma et al. have converted the attacker's problem into an equivalent convex optimization problem:
\begin{equation}\label{eq:attack_Q_optimization}
\begin{aligned}
\min_{\mathbf{{r}}\in\mathbb{R}^T, \hat{R}, Q\in\mathbb{R}^{|\mathcal{S} |\times |\mathcal{A}|}} &\Vert \mathbf{{r}}^0 -\mathbf{r} \Vert,\\
s.t.\ \ \ \ \ \ &\hat{R}(s,a) = \frac{1}{|T_{s,a}|} {\sum_{t\in T_{s,a}}r_t},\ {\hat{P}(s,s',a) = \frac{1}{|T_{s,a}|} \sum_{t\in T_{s,a}} \mathbf{1}_{\{s_t'=s'\}}  }\\
&Q(s,a) = \hat{R}(s,a) +\gamma \sum_{s'}\hat{P}(s,s',a)Q\left(s',\pi^\dag(s')\right), \forall s, \forall a,\\
& Q(s,\pi^\dag(s)) \geq Q(s,a) +\epsilon, \forall s \in\mathcal{S}, \forall a \notin \pi^\dag(s),
\end{aligned}
\end{equation}
where $T_{s,a} = \{ t | s_t =s, a_t =a\}$ is the time indices of all training data points for which action $a$ is taken at state $s$. By solving (\ref{eq:attack_Q_optimization}), the attack can mislead the RL agent into learning the strategy $\pi^\dag$ while minimizing his/her own cost of manipulating the reward data. Ma et al. have showed the existence of the optimal solution to (\ref{eq:attack_Q_optimization}) and provided a bound on the optimal cost that the attacker needs to pay to achieve the strategy $\pi^\dag$.

\begin{figure}[ht]
\centering
\includegraphics[width=.8 \columnwidth]{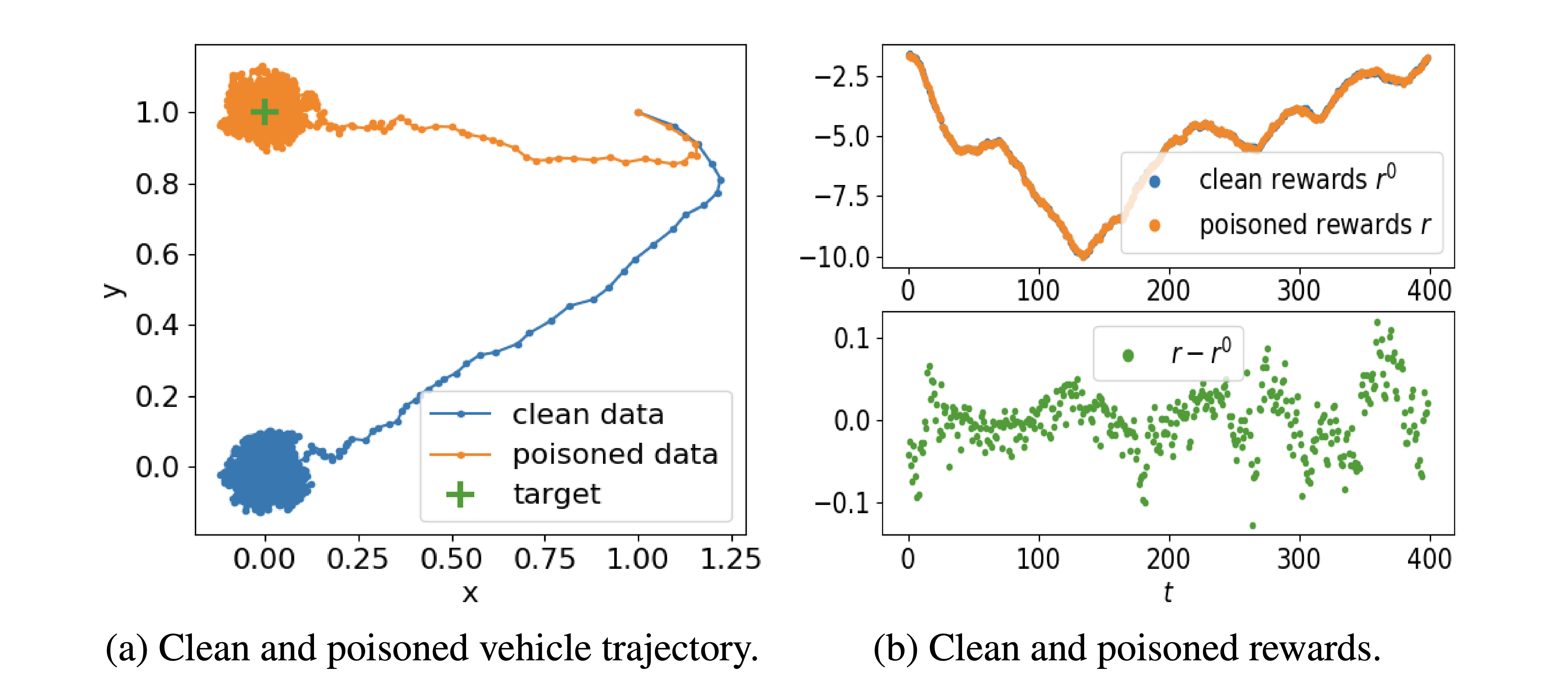} 
\caption{ 
Poisoning a vehicle running LQR in a $4$D state space.
}
\label{fig:rewardLQR}
\end{figure}

Ma et al. have applied the attacking strategy given by the solution of (\ref{eq:attack_Q_optimization}) to an RL-based linear-quadratic regulator (LQG) problem. In Fig. \ref{fig:rewardLQR} (b), the upper figure shows the true reward data and the poisoned reward data, and the bottom figure shows the difference between the two rewards. At each time step, the true reward is only poisoned slightly. However, the sight poisoning can drive the vehicle into a designated position, as we can see from Fig. \ref{fig:rewardLQR} (a). 

As the dataset becomes large $D$, i.e., $T$ increases, solving (\ref{eq:attack_Q_optimization}) becomes prohibitive. Zhang et al. have proposed an adaptive reward-poisoning attacking scheme that focuses on online RL attacks. The reward poisoning is done on the fly. Hence, their attacking strategy does not require solving a large-scale optimization problem \cite{zhang2020adaptive}. The authors demonstrated how easily an attacker could trick the agent into a designated target in a grid world problem. 

As the topic is an emerging area, most works have focused on discussing the vulnerabilities of RL instead of designing defensive mechanisms. Very few works have been studied the defense against reward poisoning attacks \cite{banihashem2021defense}. In \cite{banihashem2021defense}, Banihashem et al. have designed a defense approach to make sure the RL agent is robust against reward poisoning attacks.

\subsection{Attacks on the Sensing}

Attacks that target the sensing components, including sensors and the communication channels in CPS control systems, have been well-studied \cite{mo2010false,feng2017resilient, yuan2013resilient,huang2020cross,cardenas2008secure}. RL-based systems also share the same vulnerabilities with CPS systems --- the observation may be delayed, inaccurate, and/or missing. One difference between sensing attacks on RL and CPS is that sensing attacks can happen during the training phase, testing phase, and/or execution phase. Sensing attacks during the training phase can render the RL agent learn a `bad' policy that leads to dangerous behavior in the testing phase. Attacks in the testing phase can provide false performance evaluation to the RL agent. Attacks during the execution phase are more similar to attacks in CPS control systems, in which controllers/agents receive real-time measurements and send real-time commands. Hence, during the execution phase, any delay, inaccuracy, or absence of the observation can cause malfunctioning of the system. 

Offline RL can tolerate a certain degree of delay of the feedback signals, including reward signals, state signals, since offline RL starts after gathering a collection of data points. However, online RL can be sensitive to delays. In 2003, \cite{katsikopoulos2003markov} considered delayed state observation, in which the state observation is delayed, and the RL agent needs to select an action without any knowledge of its current state. For a fixed delay $d$, the authors formed a non-delayed MDP whose size grows exponentially in $d$. In \cite{lancewicki2020learning}, Lancewichi et al. have studied the delayed feedback in online reinforcement learning algorithms. In their work, an attacker can strategically choose the delay $d^t$, meaning that the cost signal and the state observations are only available at the end of episode $t + d^t $. The authors proposed a policy optimization-based algorithm that can achieve near-optimal high-probability regret of order $\tilde{O}(\sqrt{T}+\sqrt{D})$, where $T$ is the number of episodes and $D =\sum_{t} d^t$ the total delay. More recently, Ramstedt et al. \cite{bouteiller2020reinforcement} proposed the delay-correcting actor-critic algorithm that can achieve better performance in environments with delays.

RL can be an easy target of DoS/jamming attacks and spoofing attacks and can suffer a significant loss if the attacks are successfully launched. Take the autonomous vehicle as an example. If the state is the location of the RL agent measured by GPS coordinates, spoofing of GPS signals by the attacker can lead to incorrect navigation policy and hence, a collision. Although RL suffers a high risk and the potential loss of being attacked,  very few works have focused on RL with missing or inaccurate observations from a security perspective. Hence, the attack models and the defense mechanisms about DoS attacks and spoofing attacks on RL sensing remain open for future studies. 

RL agents observe the environment via sensors. Under DoS attacks, the state observations cannot reach the RL agent, and the agent becomes unmindful of the states. For remote sensing, jamming attacks can be launched on the communication channels. For in-house sensing, an example is sequential blinding of the cameras in an RL-based autonomous vehicle via lasers, leading to a collision. On behalf of the attack, researchers can address the following interesting questions: when and where to launch DoS/jamming attacks to create the most damage to the RL agent, whether the attacker can trick the RL agent into a nefarious policy only through jamming attacks, etc.; on behalf of the agent, researchers can design a resilient RL algorithm that is less sensitive to missing state observations \cite{feng2017resilient}, employ prediction methods when state observations are missing \cite{huang2019continuous}, or study the effect of DoS attacks on existing RL algorithms. Attacks can also manipulate the sensing signals by spoofing attacks. The problem for the attacker can be to design a manipulation strategy. The manipulation strategy can be characterized by a map $K$ that maps a state $s$ into a probability simplex over all possible states $\Delta\mathcal{S} = \{p^o(s),s\in\mathcal{S}\}$. For example, $\Delta\mathcal{S} = K(s_t)$ indicates that given the true state $s_t$ at time $t$, the RL agent will instead observe the state $s_t'\in\mathcal{S}$ with probability $p^o(s'_t)$. The map $K$ is similar to the concept of observation kernel in partially observable Markov decision process \cite{krishnamurthy2016partially} except that now the map $K$ is crafted by the attacker to mislead the RL agent and achieve his/her malicious purposes.

One concept that is close yet different to sensing attacks on RL is environment poisoning attacks, which is the focus of many works regarding the security of RL \cite{xu2021transferable,sun2020vulnerability,rakhsha2020policy}. In environment poisoning attacks, the attacker can perturb the environment, i.e., the transition probability in the MDP. As is demonstrated by Behzadan and Munir in \cite{behzadan2019adversarial}, through sequential reconfiguration of obstacles on the road, an attacker can manipulate the trajectory of an RL-based autonomous vehicle in the testing phase. In 2020, Rakhsha et al. \cite{rakhsha2020policy} study the problem of how to teach a policy to the RL agent through transition dynamics poisoning. We describe the attack model proposed by the authors using the following description:
\begin{itemize}
    \item \textit{Information the attacker knows}: The attacker is omniscient who knows almost all information including the original MDP.
    \item \textit{Actions available to the attacker:} The attacker can turn the original transition dynamics denoted by $P$ into the manipulated transition dynamics $\tilde{P}$.
    \item \textit{The objective of the attacker:} The objective of the attacker is to mislead the RL agent into a nefarious policy $\pi^\dag$ while minimizing his/her cost of poisoning transition dynamics, i.e.,
    $$
    \min_{\tilde{P}} \Vert P - \tilde{P}\Vert_\rho \coloneqq \left(\sum_{s,a}\left( \sum_{s'} |p(s,s',a) - \tilde{p}(s,s',a)| \right)^\rho\right)^{1/\rho}.
    $$
\end{itemize}

The attacker can find the optimal way of manipulating the transition dynamics by solving the following optimization problem:
\begin{equation}\label{eq:environment_pois_opt}
\begin{aligned}
\min_{\tilde{P}, \mu^{\pi^\dag}, \mu^{\pi^\dag \{s;a\}} }\ \ \ \ &\Vert P - \tilde{P}\Vert_\rho\\
s.t.\ \ \ &\textrm{$\mu^{\pi^\dag}$ and $P$ satisfy (\ref{eq:kolgorov_forw})}\\
&\forall s,a \neq \pi^\dag(s):\mu^{\pi^\dag\{s;a\}}\textrm{ and $P$ sastisfy (\ref{eq:kolgorov_forw})},\\
&\sum_{s'} \mu^{\pi^\dag}(s')\cdot r(s',\pi^\dag(s')) \geq \sum_{s'} \mu^{\pi^\dag\{s,a\}}(s') \cdot r(s',\pi^{s,a}(s')) + \epsilon,\\
&\forall s,a,s': \tilde{P}(s,s',a) \geq \delta \cdot P(s,a,s').
\end{aligned}
\end{equation}
Here, $\pi\{s,a\}$ is a neighbor policy of the policy $\pi$ defined as
$$
\pi\{s,a\}(s') = \begin{cases}
  \pi(s')\ \ \ &s'\neq s,\\
  a\ \ \ &s' =s.
\end{cases}
$$
The neighbor policy $\pi\{s,a\}$ is almost the same with policy $\pi$ except that at state $s$, the neighbor policy points to action $a$. Here, $\mu^\pi$is the stationary distribution given the transition probability $P$ and policy $\pi$, which satisfies
\begin{equation}\label{eq:kolgorov_forw}
    \mu^\pi(s) = \sum_{s'} p(s',s,\pi(s'))\cdot \mu^\pi(s').
\end{equation}
In the optimization problem (\ref{eq:environment_pois_opt}), the first three constraints are the conditions needed to be satisfied to make sure the agent learns the policy $\pi^\dag$. The last constraint specifies how much the attacker can decrease the original values of transition probabilities, where $\delta \in (0,1]$. The authors later demonstrated that using the attacking rule obtained from (\ref{eq:environment_pois_opt}), the attacker can achieve his objective by only altering a small number of transition probabilities \cite{rakhsha2020policy}. However, this approach faces scalability issues and the assumption that the attacker knows the MDP is a stringent one, which makes the attack a white-box attack. In 2021, Xu et al. \cite{xu2021transferable} studies environment-dynamics poisoning attacks at training time which expands the attacks to black-box attacks. The authors propose an attack model that prompts the momentary policy of the agent to change in the desired manner with minimal dynamics manipulation. More recently, Sun et al. \cite{sun2020vulnerability} proposed an attack model that does not require any knowledge of the MDP.  The attack model follows the so-called vulnerability-aware adversarial critic poisoning strategy. The authors demonstrated that the attacking strategy successfully prevents agents from learning a good policy or mislead the agents to a target policy with low attacking effort.

\subsection{Attacks on the Actuator}
Many works have been done to study attacks on the actuators or the communication channel between the actuators and the controllers for CPS control systems \cite{teixeira2012attack,huang2018reliable,fawzi2014secure,an2018lq,wu2007design,gupta2010optimal}. However, there are very few studies investigating attacks targeting the RL agent's actuators. RL agents influence their environments by performing actions $a_t$ via actuators. To ensure that the action is timely and accurately applied to the environment, the RL agent must guarantee first that the action command is timely and accurately transmitted to the actuator in the field. Second, the actuator executes the received commands in a timely and accurate manner. However, the attacker may exploit these vulnerabilities. Suppose the attacker can launch jamming attacks on the communication channel or manipulate the actuator. In that case, the actual action performed will be different from the one chosen by the RL agent, and hence the observed experience is corrupted. The learning results will also be corrupted if the RL agent learns from the corrupted experience without noticing the attacks.
Future works need to focus on investigating how attacks on the actuator affect the performance of the RL agent in training, testing, and implementing phases.
 
\subsection{Discussion and Future Outlooks}
While there is vast literature on IoT and CPS security issues, little is known about the security issues of RL. However, security issues of RL exist and are getting increasingly significant as the application domain of RL widens. This section reviews the literature that studies RL under an adversarial environment, where three vulnerabilities are targeted by the attacker: the reward signals, the sensing, and the actuating. These are the three signals that need to be exchanged timely and accurately between the agent and its environment, which becomes the vulnerabilities of RL-enable systems that the attacker can exploit. Unlike real-time systems such as CPS control systems, attacks on different phases of RL may serve different purposes. Attacks during the training phase can render the RL agent fail to learn a `good' policy. Attacks in the testing phase can provide false performance evaluation to the RL agent. Attacks during the execution phase are more similar to attacks in CPS control systems, in which controllers/agents receive real-time measurements and send real-time commands. Hence, during execution phase, any delay, inaccuracy, or absence of the observation can cause malfunctioning of the system.

Within relatively small literature on the security of RL, most works have focused on the attacks on reward signals, and few papers have investigated the poisoning attacks on the transition dynamics. They mainly focus on deceptive attacks that falsify the values of either the reward signals or the transition probabilities. To the best of our knowledge, very few works are dedicated to DoS/jamming attacks on either the reward, the sensing, or the actuating. Hence, we believe there is an emerging research opportunity since DoS/jamming attacks are considered as the most common attacks that could happen to a networked system \cite{cardenas2008secure}. Another gap that needs to be narrowed is the attacks on sensing and actuating for RL-enabled systems. We notice that few works have only addressed the problem of missing or noised sensing/actuating signals from a non-security perspective that achieves robustness for the RL-enabled system. Only with a good understanding of how different attacks affect the RL, one can further safeguard our RL-enabled systems.

\section{Conclusions and Discussions}\label{sec:conclusions}
Cyber-resilient mechanisms (CRM) complement the imperfect cyber protections by strategically responding to unknown threats and maintaining the critical functions and performances in the event of successful attacks. A CRM can be viewed as a feedback system that consists of three critical components: sensing, reasoning, and actuation. 
Sensing aims to acquire information about the system as well as the footprint of the attacker. Reasoning builds on the acquired information to infer the attack behaviors and the design of the optimal resilience strategies. Actuation reconfigures the system according to the optimal strategy by adapting the system parameters and attributes to unknown threats.  The sensing-reasoning-actuating feedback loop establishes an adaptive and dynamic system architecture for cyber resilience.


Reinforcement learning (RL) is a suitable data-driven framework for cyber resilience which implements the feedback-system architecture. The update of the system reconfiguration strategies dispenses with the accurate system models and relies on only the observations of the system state and its associated cost at each iteration. 
We have classified the design of RL-based CRM (RL-CRM) based on the type of vulnerabilities the system aims to mitigate. 
We have demonstrated that the RL-CRM has enabled an adaptive and autonomous way to configure moving target defense, engage attackers for reconnaissance, and guide human attention to mitigate visual vulnerabilities.  
We have learned from the literature that posture-related defense technologies are mature, while the mitigation solutions for information-related and human-induced vulnerabilities are still underdeveloped. 
New advances are needed to detect and deter deceptive attacks. Besides compensating for information disadvantage, the defender can proactively create uncertainties and increase the attack cost by designing defensive deception. To defend against human-induced vulnerabilities, we have observed the need for new designs in human-machine interactions.  Besides indirectly designing corrective compensation for human-induced vulnerabilities, we also need to focus on (non)monetary incentives supported by human-study evidence and human-assistive technologies to elicit desirable human behaviors. 

We have discussed the RL solutions to mitigate vulnerabilities of the cyber systems and the vulnerabilities of the RL itself. Vulnerabilities of RL exist and are growing increasingly significant as the application domain of RL broadens. The RL agent exchanges three kinds of time-sensitive signals with the environment, including the rewards, the state observations, and the action commands.  The three signals become the vulnerabilities of RL-enable systems that the attacker can influence. We have reviewed the current state-of-the-art involving the attacks on the three kinds of signals. These works show that under proper conditions, a small malicious perturbation of the feedback signals can mislead the RL agent to learn any policy desired by the attacker. These results serve as a warning about the seriousness of the security issues of RL-enabled systems.

The study of vulnerabilities of RL from a security point of view is an emerging topic. We have seen a surge of publication on the topic in the last two years. Most works have focused on the attacks on reward signals, and a few papers have investigated the poisoning attacks on the transition dynamics. However, almost no works have studied the attacks on sensing and actuating for RL-enable systems, which is a considerable research gap that needs to be filled. Furthermore, current literature primarily focuses on deceptive attacks that falsify the values of either the reward signals or the transition probabilities. Since DoS/jamming attacks are considered the most common attacks that could happen to a networked system, more works should be dedicated to DoS/jamming attacks on either the rewards, the state observations, or the actions commands.

\subsection{Feedback Architectures}
The RL-CRMs presented in this work follow the standard feedback architecture. This architecture can be enriched and extended to several more sophisticated ones. One is the nested feedback loops, where one RL feedback loop is coupled in another RL feedback loop. This architecture is useful to separate and then fuse the learning of distinct system components of the cyber system. For example, one RL feedback loop is used to acquire the attack footprint and learn its intent and capabilities. In contrast, the other feedback loop is used to acquire information regarding its system state. The two feedback loops can be fused for making online defense decisions in response to an unknown threat. 

Another architecture is a mixture of feedback and open-loop structures. Leveraging the ideas from moving-horizon control and estimation, the CRM can make a moving-horizon plan by looking $N$ stages into the future and optimizing the cyber resilience for the current stage and $N$ stages-to-go. This approach would require an open-loop prediction of the system under the attack and feedback-driven sensing of the environment and reasoning of the optimal moving-horizon resilience strategies. This architecture enables a look-ahead strategy that can prepare for a sequence of forthcoming events and improve resilience.

A centralized CRM would not be sufficient for large-scale cyber systems to provide a timely response since the amount of data would increase exponentially with the size, and so does its processing time.  A distributed CRM can locally monitor a subsystem and coordinate with other CRMs to achieve a global CRM performance. In doing so, we can achieve local cyber resilience for the subsystem as well as the global resilience of the entire system. When one subsystem is attacked and fails, the remaining subsystems can adapt to it, maintain their local functions, and recover from the failure of one subsystem. This decentralized mechanism aligns with the recent advances in multi-agent control systems \cite{chen2019control}, distributed control theory \cite{bakule2008decentralized}, and game theory \cite{huang2020dynamic,zhu2015game}. They can provide a theoretical underpinning for the design of distributed cyber-resilient systems.

\subsection{Reinforcement Learning}
From the three applications of RL-CRM designs, we have identified several research challenges to be addressed in the future works in this direction. First, it is important to deal with system and performance constraints in the learning process. Cyber systems have many system constraints that need to be taken into account explicitly. For example, certain addresses or functions that are not allowed or undesirable when configuring the moving target defense. The performance of the cyber systems can impact the performance of the physical systems that they serve. Hence, the requirement on the physical system performance naturally imposes a constraint on the performance of the cyber systems. Hence, the CRM would need to adapt and respond while satisfying these performance constraints.

A second challenge is to improve the learning speed. The goal of CRM is to restore the cyber system after an attack. Fast learning would enable a speedier and more resilient response to the attack. To achieve it, we would need to resort to control-theoretic ideas, such as optimal control \cite{kirk2004optimal} and adaptive control theory \cite{aastrom1983theory},  and leverage recent advances in reinforcement learning to speed up the convergence rate or improve the finite-time learning performances. A third challenge is to deal with the nonstationarity of the cyber systems. The classical RL algorithms assume that the environment is stationary and ergodic. In many cybersecurity applications, the systems are nonstationary. The system parameters and attributes change over time. For example, the attack surface may grow when the system is connected with other nodes or used by new users. There is a need to develop nonstationary RL schemes for cyber systems that guarantee performance in a finite horizon. 





\subsection{Emerging Applications}

Beyond the applications presented in this work, there are many emerging ones where RL-CRM can play a pivotal role in improving their security stature. One crucial area is the cyber resilience for  wireless communication systems under jamming and spoofing attacks. Wireless devices can prepare for a jammer with limited power by introducing redundant links and channels \cite{zhu2011eavesdropping,xu2017game}. An RL-CRM can learn from the communication and the environment to reconfigure adaptively their routes, channels, or access points when they are subject to jamming attacks \cite{zhu2012interference,zhu2011dynamic}. The devices can also leverage moving target defense to create an RL-CRM \cite{clark2012deceptive} to attract jammers to attack a fake route in order to protect the legitimate communication.


The concept of cyber resilience is also applicable to safeguard Cyber-Physical Systems (CPS) and the Internet-of-Things (IoT) networks. The impact of the successful attacks on CPS and IoT systems is not limited to cyber performance. They can create catastrophic consequences such as the meltdown of nuclear energy systems \cite{zhao2021game,zhao2020finite}, the blackout of the electric power grid \cite{zhu2012game,maharjan2013dependable}, the disruption in gas pipelines \cite{bajpai2007securing}, and casualties in road accidents \cite{sheehan2019connected}. The CPS resilience deal with resilience at both cyber and physical layers of the system. The physical-layer resilience takes into the physical-layer performances and requires a learning-based control system to respond and restore the degrading performance resulting from successful attacks. Cyber resilience and physical resilience are interdependent. An effective CRM would restore the cyber performance and thus reduce the impact of the threat on the physical layer of the system, therefore facilitating physical resilience. There is a need to create a cross-layer resilience design framework \cite{zhu2020cross} to provide a holistic view toward CPS resilience and reduce cross-layer cascading failures \cite{huang2017large}. 

Cyber insurance is an alternative paradigm to improve cyber resilience from a non-technical and economic perspective. Cyber insurance aims to transfer the unpreventable and unmitigable risks to a third party to reduce the economic impact of cyber attacks on an organization. As the attacks are becoming financially motivated, e.g., the rise of ransomware, cyber insurance is an important mechanism that needs to be integrated into the RL-CRM to mitigate further the economic loss induced by the loss of cyber performance. Interested readers can refer to \cite{hayel2015attack,zhang2019mathtt,zhang2017bi} for the discussions on cyber insurance as a tool for cyber risk management.



\bibliographystyle{elsarticle-num}
\bibliography{ARC}







\end{document}